\documentclass{article}
\usepackage{amsmath,amsfonts}
\usepackage{algorithmic}
\usepackage{algorithm}
\usepackage{array}
\usepackage[caption=false,font=normalsize,labelfont=sf,textfont=sf]{subfig}
\usepackage{textcomp}
\usepackage{stfloats}
\usepackage{booktabs}
\usepackage{verbatim}
\usepackage{graphicx}
\usepackage[most]{tcolorbox}
\usepackage{paralist}
\usepackage{xspace}
\usepackage{multirow}
\usepackage{pythonhighlight}
\usepackage{hyperref}
\hypersetup{
colorlinks = true, 
linkcolor=black, 
citecolor=black, 
urlcolor=black 
}
\usepackage{xcolor}

\newcommand{\targetState}{\ensuremath{\mathit{t}}\xspace}
\newcommand{\method}{\ensuremath{\mathit{QOIN}}\xspace}

\newcommand{\pof}[1]{\ensuremath{\mathit{POF_{#1}}}\xspace}
\newcommand{\pos}[1]{\ensuremath{\mathit{POS_{#1}}}\xspace}
\newcommand{\odr}[1]{\ensuremath{\mathit{ODR_{#1}}}\xspace}
\newcommand{\pofName}{POF\xspace}
\newcommand{\posName}{POS\xspace}
\newcommand{\odrName}{ODR\xspace}
\newcommand{\frequency}[1]{\ensuremath{\mathit{frequency_{#1}}}\xspace}
\newcommand{\hellDist}[2]{\ensuremath{\mathit{HL_{#1-#2}}}\xspace}
\newcommand{\ideal}{Ideal\xspace}

\newcommand{\X}{X\xspace}
\newcommand{\Atwelve}{\ensuremath{\hat{A}}\textsubscript{12}\xspace}

\newcommand{\jsd}{\ensuremath{\mathit{JSD}}\xspace}
\newcommand{\revision}[1]{\textcolor{black}{#1}}

\usepackage{PRIMEarxiv}

\title{Mitigating Noise in Quantum Software Testing Using Machine Learning}

\author{
  Asmar~Muqeet\\
  Simula Research Laboratory \\
  University of Oslo \\
  Oslo\\
  \texttt{asmar@simula.no} \\
   \And
  Tao~Yue \\
  Simula Research Laboratory \\
  Oslo\\
  \texttt{taoyue@gmail.com} \\
   \And
  Shaukat~Ali \\
  Simula Research Laboratory and \\
  Oslo Metropolitan University \\
  Oslo\\
  \texttt{shaukat@simula.no} \\
   \And
  Paolo~Arcaini \\
  National Institute of Informatics \\
  Tokyo\\
  \texttt{arcaini@nii.ac.jp} \\
}

\begin{document}
\maketitle

\begin{abstract}
Quantum Computing (QC) promises computational speedup over classic computing for solving complex problems. However, noise exists in current and near-term quantum computers. Quantum software testing (for gaining confidence in quantum software's correctness) is inevitably impacted by noise, to the extent that it is impossible to know if a test case failed due to noise or real faults. Existing testing techniques test quantum programs without considering noise, i.e., by executing tests on ideal quantum computer simulators. Consequently, they are not directly applicable to testing quantum software on real quantum computers or noisy simulators. To this end, we propose a noise-aware approach (named \method) to alleviate the noise effect on test results of quantum programs. \method employs machine learning techniques (e.g., transfer learning) to learn the noise effect of a quantum computer and filter it from a quantum program's outputs. Such filtered outputs are then used as the input to perform test case assessments (determining the passing or failing of a test case execution against a test oracle). We evaluated \method on IBM's 23 noise models, Google's two available noise models, and Rigetti's Quantum Virtual Machine (QVM), with nine real-world quantum programs and 1000 artificial quantum programs. We also generated faulty versions of these programs to check if a failing test case execution can be determined under noise. Results show that \method can reduce the noise effect by more than $80\%$ on the majority of noise models. To check \method's effectiveness for quantum software testing, we used an existing test oracle for quantum software testing. The results showed that \method attained scores of $99\%$, $75\%$, and $86\%$ for precision, recall, and F1-score, respectively, for the test oracle across six real-world programs. For artificial programs, \method achieved scores of $93\%$, $79\%$, and $86\%$ for precision, recall, and F1-score. This highlights \method's effectiveness in learning noise patterns for noise-aware quantum software testing.
\end{abstract}

\keywords{Software~and~its~engineering \and Software~testing~and~debugging \and Computing~methodologies \and Instance-based~learning \and Quantum~Computing \and Machine~learning.}

\section{Introduction}\label{introduction}

There has been an increased interest in quantum software engineering over the past few years, focusing on designing, developing, and testing quantum computing (QC) applications~\cite{QSE_survey,Serrano2022bookQSE,search,Long2022, mutation,Rui_mutation,bug1,bug2}. This growth of interest is due to the computational power promised by quantum computers to solve a particular class of problems more efficiently than classic computers~\cite{speed}. In addition, quantum computers (IBM~\cite{ibm}, Google~\cite{google}, Rigetti~\cite{Rigetti}), and quantum computer simulators such as QuEST~\cite{QuEST}, QX~\cite{QX}, and IBM's Qiskit Aer simulator~\cite{qiskit} are becoming available. 
However, quantum computers are susceptible to hardware noise due to immature hardware and environmental factors (e.g., magnetic fields, radiations)~\cite{Chuang1995, noise_benchmark1}. Noise affects the accuracy of calculations a quantum computer performs, thus resulting in incorrect program outputs. Such computers with inherent noise are known as Noisy Intermediate-Scale Quantum (NISQ) computers~\cite{NISQ}.

Quantum software testing aims to cost-effectively find quantum software bugs to achieve a certain level of confidence in their correctness~\cite{ontest,Garcia2023,MiranskyyICSE2020,CACM,QSE_survey}. Testing quantum software is challenging due to the inherent quantum mechanics' features, such as superposition and entanglement~\cite{ontest,CACM}. In addition, noise brings another layer of complexity to the challenge. For example, when checking test results, it becomes difficult to conclude whether a test case execution failed due to a faulty quantum program or noise. Existing quantum software testing approaches mainly focus on adopting classic software testing techniques, such as applying combinatorial testing~\cite{Combinatorial}, defining coverage criteria~\cite{coverage,Long2022,quraTestASE23}, relying on search-based testing~\cite{search,mutation2}, mutation testing~\cite{mutation1,Fortunato2022,mutation2,mutation}, metamorphic testing~\cite{Rui_metamorphic,morphq}, property-based testing~\cite{property}, and fuzz testing~\cite{fuzz}. However, these works perform testing on ideal (i.e., noise-free) quantum computer simulators, so test results cannot be trusted when testing on quantum computers with noise or noisy simulators.
In addition, using simulators for test execution is time-consuming. For instance, for a 7-qubit quantum program, it may take 12 hours to complete an execution of a single input~\cite{simulators_are_slow}. Testing usually requires the execution of multiple inputs; hence, it is limited by the computational cost of simulators of a few qubits. Therefore, we need proper ways to filter the noise to support the testing of quantum programs directly on NISQ computers. This is, however, not easy since each NISQ computer exhibits a different noise effect on the outputs of the same program~\cite{Martina2022}, implying that a program's outputs (due to noise) vary on each NISQ computer.

In summary, on NISQ computers, quantum noise prevents us from applying existing testing methods for testing quantum programs directly. Therefore, we propose an approach to make {\it Quantum sOftware testIng Noise-aware} (\method) to alleviate the effect of noise on testing quantum programs on NISQ computers. 
\method uses supervised machine learning to reduce the effect of noise on test results. \method trains a fully connected neural network to learn general noise patterns of a NISQ quantum computer. The effect of noise on the output of a quantum program is not only specific to each NISQ computer but also specific to each program. Therefore, a network that learns only the general noise pattern of a NISQ computer may not be able to minimize circuit-specific noise. To alleviate this limitation, \method uses a transfer learning method to learn circuit-specific noise. \method uses circuit-specific models to predict the correct output of a program from a noisy output. Such predicted output can then be assessed with a given test oracle.

Our key contributions are:
\begin{inparaenum}[1)]
\item A machine-learning based approach to reduce noise's effect on the test results of a quantum program;
\revision{\item Empirical evaluation using IBM's 23 quantum computer noise models, Google's two available noise models and Rigetti's Quantum virtual machine (QVM), with nine open-source real-world quantum programs;}
\item An extended empirical evaluation with 1000 diverse quantum circuits, generated with Qiskit to assess \method's applicability for diverse circuits; and
\item Empirical evaluation of \method with a published quantum test oracle to show its effectiveness in aiding test case assessment on noisy quantum computers and simulators.
\end{inparaenum}

Our results showed that \method can effectively learn and reduce the effects of noise more than $80\%$ on the majority of backends. We used an existing test oracle~\cite{Combinatorial, search} for quantum software testing with \method. \revision{The results indicate that \method attained scores of $99\%$, $75\%$, and $85\%$ for precision, recall, and F1-score, respectively, for the test oracle across six real-world programs. In the case of 800 diverse artificial programs, \method achieved scores of $93\%$, $79\%$, and $85\%$ for precision, recall, and F1-score.}

The paper is structured as follows. Sect.~\ref{sec:background} provides the necessary background. Sect.~\ref{sec:related} positions our work in the literature. Our approach \method is presented in Sect.~\ref{sec:approach}. Sect.~\ref{sec:experimentDesign} presents the experiment design. Sect.~\ref{sec:results} provides the experiment results. Sect.~\ref{sec:overalldiscussion} discusses the overall results. Sect.~\ref{sec:conclusion} concludes the paper and discusses future directions.

\textbf{\textit{Open science.}} Our implementation and all experimental results are being made freely available~\cite{sourcecode}.

\section{Background}\label{background} \label{sec:background}
\subsection{QC Basics and Example} \label{subsec:QC}
Classic computing uses classic bits that can only be in one of two states: 0 or 1. QC's basic unit of information is quantum bit or \textit{qubit}, which can exist in a \textit{superposition} of $|0\rangle$ and $|1\rangle$ with amplitudes $(\alpha)$ associated with them. Amplitude $\alpha$ is a complex number consisting of a \textit{magnitude} and a \textit{phase} in its polar form. We represent a qubit in the Dirac notation~\cite{dirac} as: $|\psi\rangle = \alpha_0 |0\rangle + \alpha_1 |1\rangle$, where $\alpha_0$ and $\alpha_1$ are the amplitudes associated with states $|0\rangle$ and $|1\rangle$, respectively. The probabilities of a qubit being in $|0\rangle$ or $|1\rangle$ are given by the square of the magnitude of $\alpha_0$ and $\alpha_1$, with the sum of all squared magnitudes being equal to 1: $|\alpha_0|^2 + |\alpha_1|^2 = 1$. For example, the probability of being in state $|0\rangle$ is $|\alpha_0|^2$. 

Qubits are manipulated via quantum gates. A quantum gate is a unitary operator which changes a qubit's state based on a unitary matrix~\cite{basic}. For example, the \textit{Hadamard} gate puts a qubit into superposition. Currently, we program gate-based quantum computers as quantum circuits in which the logic of a program is represented as a sequence of quantum gates applied on qubits.

Fig.~\ref{fig:ghzcode} shows a GHZ (Greenberger-Horne-Zeilinger) state, defined with a three-qubit \textit{entanglement} program written in Qiskit~\cite{qiskit}.
\begin{figure}[!tb]
\centering
\begin{python}[numbers=left,stepnumber=1,numberfirstline=true]
 1. #initialize the empty circuit
 2. qc = QuantumCircuit()
 3. # create 3 qubits
 4. qubit_1 = QuantumRegister(1,'qubit_1')
 5. qubit_2 = QuantumRegister(1,'qubit_2')
 6. qubit_3 = QuantumRegister(1,'qubit_3')
 7. # create 3 classic registers
 8. qubit_1c = ClassicalRegister(1, 'qubit_1c')
 9. qubit_2c = ClassicalRegister(1, 'qubit_2c')
10. qubit_3c = ClassicalRegister(1, 'qubit_3c')
11. # Apply a Hadamard gate on qubit_1
12. qc.h(qubit_1)
13. # Entangle qubit_1 with 2, 2 with 3
14. qc.cx(qubit_1,qubit_2)
15. qc.cx(qubit_2,qubit_3)
16. # measure the qubits for readout
17. qc.measure(qubit_1,qubit_1c)
18. qc.measure(qubit_2,qubit_2c)
19. qc.measure(qubit_3,qubit_3c)
\end{python}
\caption{Illustrating three-qubit GHZ state in Qiskit.}
\label{fig:ghzcode}
\end{figure}
The program puts three qubits in \textit{entanglement}, i.e., when being read, they have the same value (either $|0\rangle$ or $|1\rangle$). In lines 1-10, the program initializes a quantum circuit with three qubits (\textit{qubit\_1}, \textit{qubit\_2}, \textit{qubit\_3}), and three classical registers (\textit{qubit\_1c}, \textit{qubit\_2c}, and \textit{qubit\_3c}) for storing the final states of the qubits. At line 12, the program applies a \textit{Hadamard} gate on \textit{qubit\_1} to put it in the superposition of $|0\rangle$ and $|1\rangle$. At line 14, \textit{qubit\_1} and \textit{qubit\_2} are entangled via the controlled-not gate (\textit{cx}). Since \textit{qubit\_1} is already in superposition, after \textit{cx}, the program is in a superposition of $|00\rangle$ and $|11\rangle$. At line 15, the program entangles \textit{qubit\_2} and \textit{qubit\_3} via another \textit{cx}. Then, the program has all the qubits entangled: reading them will result in either $|000\rangle$ or $|111\rangle$ with 50\% probability. Lines 17-19 show that three \textit{measure} operations are applied to read the qubits and store results in classic registers \textit{qubit\_1c}, \textit{qubit\_2c}, and \textit{qubit\_3c}.

\subsection{Quantum Noise}\label{sec:background_noise}
Noise may originate from various sources. First, environmental characteristics (e.g., magnetic fields, radiations) affect computations performed by qubits~\cite{Chuang1995, noise_benchmark1}. Interactions of qubits with environments can cause disturbances and loss of information in quantum states, commonly known as \emph{decoherence}~\cite{decoherence_def}. Second, unwanted interactions of qubits exist among themselves even when perfectly isolated from the environment, called \emph{crosstalk noise}~\cite{decoherence1,decoherence2,decoherence3}; this leads to unwanted quantum states, thus affecting the computation performed by qubits. Third, noise is caused by imprecise quantum gate calibrations, which are required to optimize gate parameters to reduce errors and improve their fidelity~\cite{callibration}. Minor errors in such calibration often result in minor changes of phases, amplitude, etc., in qubits, which may not directly destroy a quantum state but could lead to undesired states after a sequence of gate operations~\cite{gatenoise}.
Note that any qubit in a quantum program can be affected by noise at any stage, resulting in an accumulated noise effect on the program's output. 

As an example, we show the impact of noise on the quantum program reported in Fig.~\ref{fig:ghzcode}; the ideal output of the program is to give almost an equal probability of having either state \(|000\rangle\) or \(|111\rangle\) after 1024 executions (see Table~\ref{tab:ghzoutput}).
\begin{table}[!tb]
\caption{The ideal and noisy outputs of a three-qubit entanglement program (see Fig.~\ref{fig:ghzcode}) after being executing 1024 times on both an ideal and a noisy simulator. Column \textit{Probability} shows the probability of having a specific output. 
}
\label{tab:ghzoutput}
\centering
\setlength{\tabcolsep}{4.5pt}
\begin{tabular}{c|cccccccc}
\toprule
\textbf{} & \multicolumn{8}{c}{\textbf{Probability}} \\
\midrule
\textbf{State} & \textbf{000} & \textbf{001} & \textbf{010} & \textbf{011} & \textbf{100} & \textbf{101} & \textbf{110} & \textbf{111} \\
\midrule
\textbf{Ideal} & 0.5 & - & - & - & - & - & - & 0.5 \\
\textbf{Noisy} & 0.476 & 0.013 & 0.007 & 0.016 & 0.008 & 0.019 & 0.020 & 0.443 \\
\bottomrule
\end{tabular}
\end{table}
However, when running on a NISQ computer, the program can have eight output states of various probabilities (see Table~\ref{tab:ghzoutput}). Thus, due to noise, a program can produce wrong output states or correct output states with wrong probabilities. Furthermore, each quantum computer exhibits a unique noise effect~\cite{Martina2022}; thus, outputs of the same program on different computers can be different. Moreover, noise in NISQ computers not only differs in each NISQ computer, but it is also specific for each quantum program. This requires any noise learning strategy to be not only computer-specific but also program-specific, making it harder to generalize across computers and programs.

\revision{Noise also exists in the classical world, such as Internet of Things (IoT) and cyber-physical systems~\cite{sensornoise, iotnoise}. This raises the question of whether classical noise filtering or error correction techniques can be applied to QC. While some principles from classical methods, such as error correction codes derived from information theory, can indeed find applications in QC~\cite{qec1,qec2}, it is essential to acknowledge that quantum noise possesses unique characteristics (e.g., quantum entanglement, superposition, and quantum interference), which sets it apart from classical noise, making quantum noise treatment considerably more complex. In contrast, classical noise can be described with classical probability theory and stems from random fluctuations, electronic interference, thermal effects, etc.~\cite{classicalnoise}. These classical noise sources often exhibit behaviors where noise events are independent and adhere to well-understood probability distributions like Gaussian or Poisson distributions~\cite{classicalnoise}.}


\revision{
A quantum computer's \textbf{\textit{noise model}} is a mathematical model that encapsulates the error probabilities of all qubits and gates within a quantum computer, which are characterized by a set of parameters such as qubit bit-flip errors, relaxation times (T1), dephasing times (T2), and qubit cross-talk probabilities. In summary, the noise model of a quantum computer serves as a general probabilistic representation of potential errors that may occur during quantum computation. However, this model does not encapsulate information about specific error instances in a given quantum circuit and is unsuitable for implementing fine-grained noise filtering mechanisms for individual circuits. 
In our context, we define \textbf{\textit{noise pattern}} to refer to the specific behavior of quantum noise in the context of a particular quantum circuit and a quantum backend. It describes how noise manifests itself based on the combination of qubits used and the gate operations performed on a given quantum circuit. Different quantum circuits interacting with different qubits and gates on a specific quantum processor can lead to distinct noise patterns.}

\section{Related work}\label{sec:related}
\textbf{\textit{Noise in quantum computers.}} The work in~\cite{Martina2022} studies the NISQ computer noise and concludes that such noise is distinguishable across NISQ hardware with machine-learning techniques since each computer has its own noise footprint that can be learned. Specifically, the authors used a quantum circuit as a testbed and executed it on several NISQ devices. The execution data was then used to train an ML model to classify which execution belongs to which NISQ device. The model had an accuracy of 99\%, showing that each NISQ device has a unique noise fingerprint. In contrast to~\cite{Martina2022}, our goal is to use ML not to classify but to minimize the effect of noise on the output of a quantum program. Another recent study~\cite{Ruan2022} compares executions of a quantum circuit across different NISQ computers by providing a holistic picture of the quantum circuit's fidelity on different NISQ devices to identify NISQ computers with minimal noise. 

\textbf{\textit{Quantum software testing.}} Some approaches have been proposed in the literature for quantum software testing. Huang and Martonosi~\cite{Huang2019} proposed an assertion and breakpoint strategy for debugging quantum programs using a statistical test. A projection-based method was proposed in~\cite{projection} to add assertions to quantum programs at run-time. Quito~\cite{coverage,quitoASE21tool} introduces input-output coverage criteria with two test oracles. 
QMutPy~\cite{Rui_mutation} is a mutation testing tool for quantum programs built on MutPy~\cite{Mutpy}. A metamorphic testing approach was presented in~\cite{Rui_metamorphic} to test quantum programs with defined metamorphic relations. 
QSharpTester~\cite{Long2022} presents an equivalence class partitioning approach, for quantum programs of multiple subroutines, to partition input space into classes such as classical, superposition, and mixed states. QSharpCheck~\cite{property} is a property-based test strategy for testing Q\# quantum programs, which relies on their general properties being defined as a logical style of pre- and post-conditions. QuanFuzz~\cite{fuzz} is a fuzzing testing approach for quantum programs. QuSBT~\cite{search} is a search-based strategy that generates test suites with the maximum number of failing tests. MutTG~\cite{mutation2} is a search-based generator that tries to generate test suites that kill the maximum number of non-equivalent mutants with the minimum number of test cases. QuCAT~\cite{Combinatorial} applies combinatorial testing for testing quantum programs.
%
Muskit~\cite{mutation} defines mutation operators to generate mutated versions of quantum programs. In~\cite{bug1,bug2}, bug identification and classification approaches for quantum programs were presented. Bugs4Q~\cite{bug3} is a benchmark of real-world bugs in quantum programs written in Qiskit and test cases to catch similar bugs. The authors of study~\cite{MiranskyyICSE2020} discussed the applicability of classic debugging tactics (e.g., backtracking, cause elimination, and brute force) in the context of quantum programs and showed which ones can be used for identifying bugs in quantum programs. 
\revision{To ensure the reliable support of quantum software platforms (QSP) like Qiskit~\cite{qiskit} for the development of quantum programs, two approaches, QDiff~\cite{Qdiff} and MorphQ~\cite{morphq}, have been introduced. QDiff relies on differential testing, while MorphQ uses metamorphic testing strategies for evaluating QSP.}

\revision{It is important to note that testing QSP differs from testing quantum programs. Testing QSP focuses on verifying the implementation of the QSP itself, and quantum programs serve as instruments (like benchmarks) to test various QSPs. On the other hand, testing a quantum program involves assessing and validating the logic within the quantum program itself. Among the reviewed literature, only QDiff~\cite{Qdiff} considers quantum noise. However, QDiff does not test quantum programs under the influence of noise. The primary concept of QDiff is to test QSP by randomly generating logically equivalent quantum programs. It utilizes static measures such as gate error rates and T1 relaxation time to identify and exclude quantum programs that might be significantly impacted by noise from its test suite. The generated programs serve as tools to test QSP. It is important to note that QDiff does not filter noise from the quantum program output. In our particular case, we test quantum programs directly, and we create a machine-learning model using noise data and then apply it to filter out noise from the quantum program output, ensuring that test assessments can be carried out accurately.}

\revision{\textbf{Classical noise reduction.} Noise affecting computations is not exclusive to quantum computation; it also manifests in classical computation domains like IoT and cyber-physical systems~\cite{sensornoise,iotnoise}. Classical noise reduction employs methods such as noise-free signal processing~\cite{noisefree1}, adaptive noise filtering~\cite{noisefree2}, and Bayesian inference to mitigate noise in computations. However, applying these methods to QC faces challenges due to the principles of no-cloning and state collapse in quantum mechanics~\cite{nocloning}. 
In contrast, classical information can be copied precisely; hence, classical methods rely on error information and redundancy to detect errors caused by noise~\cite{classicalnoise}. Therefore, effective quantum noise handling requires quantum-specific techniques and ML-based noise reduction methods, such as using neural networks to learn noise~\cite{MLnoise}, remain applicable in both classical and quantum contexts.}


\section{Approach}\label{sec:approach}

\begin{figure}[!tb]
\centering
\includegraphics[width=0.7\columnwidth]{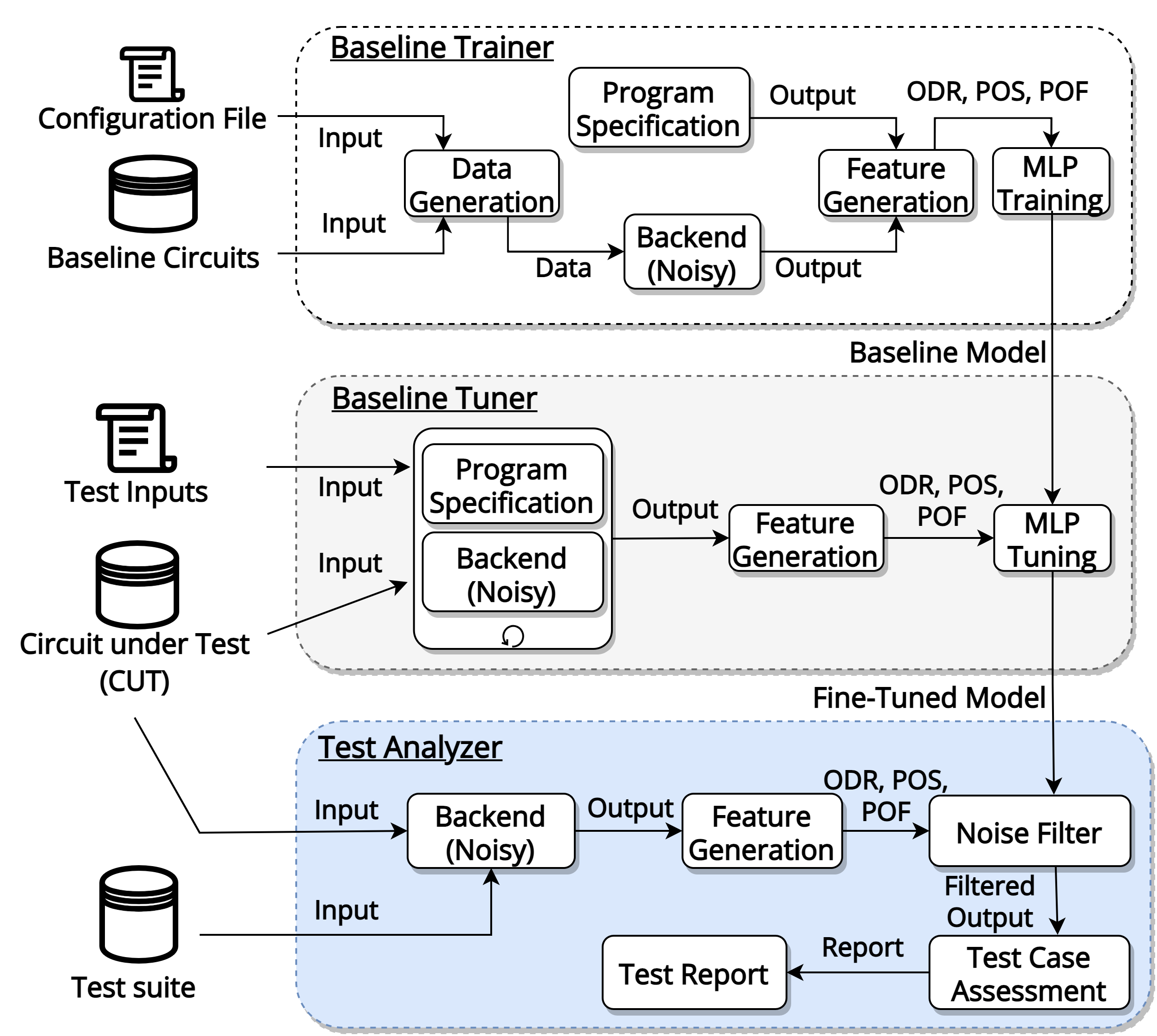}
\caption{Overview of \method. ODR is the Odds ratio for each output; POS is the Probability of success for each output; POF is the Probability of failure for each output.}
\label{fig:overview}
\end{figure}

As shown in Fig.~\ref{fig:overview}, \method has three modules: \textit{Baseline Trainer}, \textit{Baseline Tuner}, and \textit{Test Analyzer}, which will be detailed in Sects.~\ref{sec:btrainer}-\ref{sec:btest}.
%
For automated testing, a test oracle is typically required to assess the result of a test case. Such assessment is usually done by comparing an observed output of a test case against a \textit{Program Specification}. For the training purpose, \method also relies on the program specification to guide the neural network for learning the noise pattern of a NISQ computer. \method also uses a \textit{Noisy Backend}, which is either a NISQ computer or a simulator with a noise model of a NISQ computer for noisy program execution.

\subsection{Baseline Trainer}\label{sec:btrainer}

This module learns a general noise pattern of a given noisy backend since each NISQ computer has a unique noise fingerprint~\cite{Martina2022}. Therefore, we need to learn one dedicated neural network per noisy backend. \textit{Baseline Trainer} has three components: \textit{Data Generation}, \textit{Feature Generation}, and \textit{MLP Training}, described below along with a running example.

\subsubsection{Data Generation}\label{sec:data_generation}

This component takes the \textit{Configuration File} and \textit{Baseline Circuits} as input and generates the input data for quantum circuits to execute on the noisy backend. 
\textit{Baseline Circuits} are quantum circuits required to produce training and testing data for the neural network.
\textit{Configuration File} specifies the input parameters of each circuit, which are used for guiding the generation and controlling the input data required for circuit executions. A quantum circuit takes input parameters in three formats: integer, binary, and string expression. In the configuration file, one can specify each parameter's range (e.g., the minimum and maximum of an integer).
One can also specify the percentage of input space to be explored for data generation since executing a quantum circuit with all possible inputs can be computationally expensive.
For instance, as shown in the example in Fig.~\ref{fig:ghzcode}, before superposition, the input to the circuit will be binary, i.e., a three-bit string ranging from 000 to 111. In the configuration file, we can specify the percentage of these inputs we want to cover (e.g., 50\%). We can also define a possible minimum and maximum value (e.g., from 001 to 101), where the generated input will start from 001 and end at 101, depending on the percentage value.

\revision{Fig.~\ref{fig:rules} shows a snippet of the \textit{Configuration File} for three circuits.
\begin{figure}[!tb]
\centering
\includegraphics[width=0.5\columnwidth]{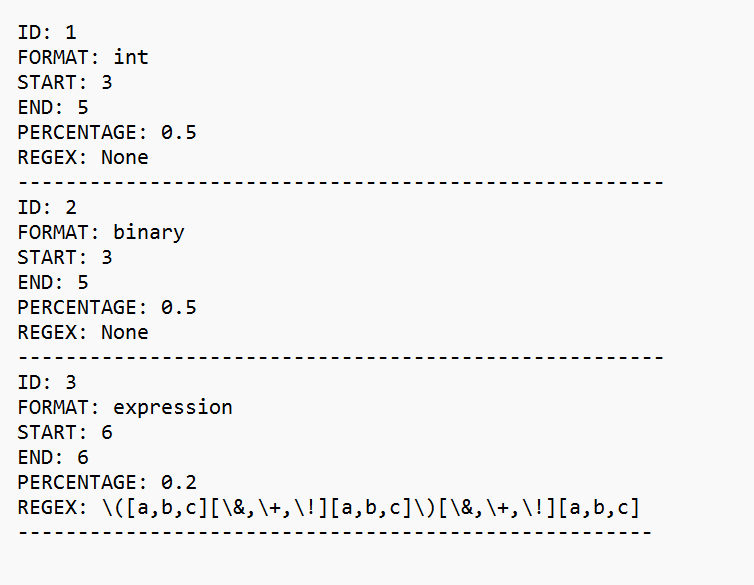}
\centering
\caption{An example configuration file for three baseline circuits.}
\label{fig:rules}
\end{figure}
\textit{ID} identifies a particular circuit in \textit{Baseline Circuits}; \textit{FORMAT} defines the input data format a circuit uses. In the case of an integer, \textit{START} and \textit{END} define the minimum and maximum integer values. All values within the minimum and maximum range are considered valid input values. For binary, \textit{START} and \textit{END} define the minimum and maximum numbers of valid bits in a binary string. Only \textit{START} is used for the expression format, and \textit{END} is ignored. \textit{START} defines the number of unique characters in a regular expression. \textit{PERCENTAGE} defines the percentage of the input space explored for data generation. Finally, for the expression format, \textit{REGEX} defines the required regular expression from which input data is generated.}

\textit{Data Generation} generates input data for executing a quantum circuit on the noisy backend (see Fig.~\ref{fig:overview}). The program specification is used as the ground truth for neural network training. The outputs from the noisy backend and the program specification are delivered to the \textit{Feature Generation} component as input. 

\subsubsection{Feature Generation}\label{sec:features}
A quantum circuit cannot be directly used as input for classic neural network training because classic neural networks require a descriptive representation of data to extract features during training~\cite{mlp, mlp2, mlp3, mlp4}. Therefore, the \textit{Feature Generation} component generates three descriptive features (i.e., \textit{Odds Ratio}, \textit{Probability of Success}, and \textit{Probability of Failure}) for each circuit in \textit{Baseline Circuits}. These features have been used to solve several ML problems~\cite{featureimportance1,featureimportance2,featureimportance3}. All these features are calculated by utilizing the outputs from both the program specification and the noisy backend. With these three features, we map the noise learning problem as a supervised regression problem, which can be addressed with a trained neural network. After the feature calculation, the three features are used as input for \textit{MLP Training}. \revision{There can be multiple states as the outcome of an execution of a quantum circuit. For example, the GHZ example (Fig.~\ref{fig:ghzcode}) has two output states: $000$ and $111$. We call them \textit{target states}. Each of the three features are calculated for each \textit{target state}.}

\revision{\textit{Probability of Success (\posName)} of a target state \targetState is the probability of the state being observed after the execution of a quantum circuit on a noisy backend, which is calculated by dividing the frequency of \targetState by the sum of frequencies of all output states:}
%
$\pos{\targetState} = \frac{\frequency{\targetState}}{\sum_{i=1}^n \frequency{i}}$.
%

\textit{Odds Ratio (\odrName)} defines the odds of one event in the presence of another one~\cite{oddsratio}. Odds ratio is calculated as the probability of one target state divided by the sum of the probabilities of the others:
%
$\odr{\targetState} = \frac{\pos{\targetState}}{1 - \pos{\targetState}}$,
%
where \pos{\targetState} defines the probability of a target state \targetState observed after the program execution on the noisy backend,
and $1-\pos{\targetState}$ is the sum of the probabilities of all the other observed states.


\textit{Probability of Failure (\pofName)} of a target state \targetState is the likelihood of no occurrence, i.e., the complement of \posName: 
\pof{\targetState} = 1 - \pos{\targetState}. 

\revision{\odrName and \pofName are features derived from \posName. The intuition behind using these three features is to capture two different scenarios in which noise can affect the program output.}

\revision{In the first scenario, where the program produces correct output states but with incorrect probabilities under noise (see Sect.~\ref{sec:background_noise}), \posName and \pofName can help the ML model capture the differences between the ideal probabilities and the probabilities under noise.}
\revision{For example, consider the program output in Table~\ref{tab:ghzoutput}. Assume the target state is $000$ with an ideal probability of $0.5$ and an observed noisy probability (\posName) of $0.47$. If the ideal value is known, correcting the observed probability involves taking the difference between the ideal and observed probabilities, in this case, $(0.5 - 0.47)$, resulting in $0.03$ as the correcting factor. However, in the absence of ground truth, the correcting factor cannot be directly calculated with only \posName. Instead, \pofName can be used as an approximate ground truth. As \pofName represents the cumulative sum of all other observed probabilities, it can be utilized to determine the correcting factor. In the example for the $000$ target state, where \posName is $0.47$ and \pofName is $0.53$ (essentially ideal probability plus some error, i.e., $0.5 + \mathit{error}$), the correcting factor can be calculated as (\pofName{} $-$ \posName), which is $(\mathit{ideal} + \mathit{error} - \posName)$. Substituting the values $(0.5 + \mathit{error} - 0.47)$ yields $0.03 + \mathit{error}$. This $\mathit{error}$ term varies for different program outputs depending on noise pattern and can be learned by ML models with sufficient data. Therefore, the first scenario of correcting the probability can be handled with \posName and \pofName for each target state.}

\revision{In the second scenario, where the quantum program can produce incorrect output states that are not part of the program specification (see Sect.~\ref{sec:background_noise}), the \odrName can be used as an input feature for ML to learn to distinguish noise-induced states from other states. \odrName may exhibit different behavior for noise-induced states compared to other states. For example, consider the program output in Table~\ref{tab:ghzoutput}. For a target state specified in the program specification, such as the $000$ state, the \odrName value is $0.9$ (calculated as $\frac{0.47}{0.53}$), which is closer to $1$, whereas, for all noise-induced states such as $001$, $010$, etc., the \odrName value is very close to the \posName value. For instance, for the state $001$, the \odrName is $0.0137$, whereas \posName for $001$ is $0.013$. This is one example of the behavior of \odrName, which holds true for quantum programs with numerous noise-induced states but lower probabilities, as illustrated in Table~\ref{tab:ghzoutput}. However, depending on the ideal output distribution of a quantum program and the effect of noise, \odrName might not exhibit the same behavior for other quantum programs. This different behavior of \odrName for noise-induced states for different programs can assist the ML model in distinguishing between noise-induced states and other states.} 

\subsubsection{MLP Training}\label{sec:mltrainer}
We chose a commonly used, fully connected, multi-layer neural network called Multilayer Perceptron (MLP)~\cite{mlp, mlp2, mlp3, mlp4} to train a neural network for our problem on the generated feature dataset. We split the dataset into training and test sets with a ratio of 80 to 20, following a common practice~\cite{split, split2, split3}. The \textit{MLP Training} component uses the Ktrain library~\cite{ktrain} for automatic neural network architecture and hyperparameter selection. The library provides standard functions that use different algorithms, such as the cyclical learning rate policy in~\cite{learningrate}, to simulate neural network training for several iterations to find optimal parameters. We selected Mean Absolute Error (MAE) as the loss function for neural network training~\cite{mae, mae2}, commonly used for training regression models. 

\subsection{Baseline Tuner}\label{sec:btuner}
As explained in Sect.~\ref{sec:background_noise}, noise has a specific fingerprint for each NISQ computer~\cite{Martina2022}. Therefore, \textit{Baseline Trainer} must learn a noise pattern for each noisy backend. However, noise also depends on the arrangement of qubits in a circuit and gate operations performed with specific inputs (named \emph{circuit-specific noise}).
For instance, quantum circuits usually have a sequence of conditional gates, such as one qubit controlled-Not and multi-qubit controlled-Not, each of which acts on a target qubit depending upon one or more control qubits. Executing a circuit with conditional gates leads to different circuit paths for each input. Conditional gates are affected by noise; therefore, each input will have a different noise effect depending upon whether the conditional gates make different execution paths in a circuit for each test input. This makes the noise also input-specific. As a result, for a given noisy backend, it is not guaranteed that the baseline MLP model learned for the backend will still be reasonably accurate when applied for all possible quantum circuits and inputs to be executed on the same backend. To address this challenge, we propose \textit{Baseline Tuner} to fine-tune the baseline MLP model trained in the \textit{Baseline Trainer}.


Specifically, \textit{Baseline Tuner} incorporates circuit-specific noise into the baseline MLP model for each noisy backend. As shown in Fig.~\ref{fig:overview}, a Circuit Under Test (CUT) and a set of predefined test inputs of the CUT are used as the input to \textit{Baseline Tuner}.
Note that, regarding the predefined test inputs, we refer to inputs used by developers to verify if a certain implementation of a program is correct, i.e., these predefined test inputs are non-failing test cases. \textit{Baseline Tuner} executes the CUT on the noisy backend with the predefined test inputs, aiming to execute the CUT on different combinations of circuit paths so that the baseline MLP model can learn the input-specific and circuit-specific noise effects on the CUT.

The \textit{Feature Generation} component of \textit{Baseline Tuner} uses the outputs from the noisy backend and the program specification to calculate the feature dataset (see Sect.~\ref{sec:features}) for the \textit{MLP Tunning} component. The \textit{MLP Tuning} component follows a transfer learning process on the baseline model. In ML, transfer learning is a method in which a pre-trained neural network on a similar task is used as a starting point for a new task~\cite{transferleraning}. In our context, using a pre-trained network (i.e., \textit{Baseline Model}) allows the \textit{Fine-Tuned Model} to learn noise patterns with fewer data. The dataset for transfer learning is generated by executing the CUT with given test inputs multiple times (e.g., 100 times, as set in our empirical study). Due to noise and inherent uncertainties of QC, the output of a CUT for each execution is different. Therefore, executing a test input multiple times allows the \textit{Fine-Tuned Model} to capture variations of output states. The hyper-parameters for the transfer learning were set automatically according to the specifications in~\cite{ktrain}.

\subsection{Test Analyzer}\label{sec:btest}
This component adds a filter layer between the output of a quantum program and the input to the test assertion module of a given quantum software testing strategy to enable a testing strategy to work on a noisy backend. 
Specifically, the \textit{Test Analyzer} component takes the \textit{CUT} and a \textit{Test suite} as input. \textit{Test suite} is a collection of test cases that, for example, has been automatically generated by a testing method (e.g., QuCAT~\cite{Combinatorial}, QuSBT~\cite{search}). The CUT and \textit{Test suite} are executed on the noisy backend. The \textit{Feature Generation} component of \textit{Test Analyzer} processes the output from the noisy backend and calculates the three features (see Sect.~\ref{sec:features}) required by the \textit{Noise Filter} component. The \textit{Noise Filter} component uses the three calculated features and the \textit{Fine-Tuned Model} from \textit{Baseline Tuner} to filter the noise out from the output of the CUT. This filtered output will then be consumed by the test assertion module of any proposed test method to determine the passing or failing of a test case against the given test specification.

\revision{As of now, \method does not introduce its own test assessment module. Instead, to demonstrate its effectiveness, we have integrated \method with test assessment modules from two previously published works~\cite{Combinatorial, search}. The integrated module provides two distinct test oracles: \textbf{Unexpected Output Failure (UOF):} This oracle examines the program's output to identify any unexpected output states not included in the program specification. \textbf{Wrong Output Distribution Failure (WODF):} This oracle evaluates the program's output distribution and compares it to the program specification using statistical tests. By incorporating these oracles, \method can assess the correctness of quantum programs and detect discrepancies between program outputs and their intended specifications.}

\section{Experiment Design}\label{sec:experimentDesign}
We aim to answer two research questions:
\begin{compactitem}
\item[\textbf{RQ1}] How effective is \method in reducing the noise effect on quantum programs' outputs?
\item[\textbf{RQ2}] How can \method help testing methods improve their test case assessment? 
\end{compactitem}

Below, we present benchmarks in Sect.~\ref{sec:benchmarks}, followed by the experiment settings in Sect.~\ref{sec:experimentSetting}, and evaluation metrics in Sect.~\ref{sec:metrics}.



\subsection{Benchmarks}\label{sec:benchmarks}

\textbf{Real-world Benchmarks.} We selected nine quantum programs from two repositories~\cite{benchmark1, benchmark2} based on the following inclusion criteria:
%
\begin{inparaenum}[1)]
\item written in Qiskit and publicly available to ensure the experiment's reproducibility,
\item input qubits greater or equal to three to ensure that the \textit{Data Generation} component can have sufficient input data space, and
\item executable on \revision{noisy backends from IBM, Google and, Rigetti}.
\end{inparaenum}
The characteristics of the real-world benchmarks, measured in the number of qubits, gate operations, and circuit depth (the longest sequence of quantum gates in a circuit), are presented in Table~\ref{table:real_program_table}.
\begin{table}[!tb]
\caption{Characteristics of the real-world benchmarks. Programs with $n$ in column \textit{\#Qubits} denote that, theoretically, they can be programmed with any number of qubits given enough hardware resources.}
\label{table:real_program_table}
\centering
\setlength{\tabcolsep}{2.6pt}
\begin{tabular}{c|ccc}
\toprule
\textbf{Real-world Benchmarks} & \textbf{\#Qubits} & \textbf{\#Gates} & \textbf{Depth} \\
\midrule
Addition & 7$\sim$n & 11 & \multicolumn{1}{c}{17} \\
Simon & 6$\sim$n & 6 & \multicolumn{1}{c}{14} \\
Greenberger–Horne–Zeilinger state (GHZ) & 3$\sim$n & 5 & \multicolumn{1}{c}{7} \\
Binary Similarity & 7 & 6 & \multicolumn{1}{c}{9} \\
N-CNOT & 7 & 6 & \multicolumn{1}{c}{9} \\
Permutations & 7 & 3 & \multicolumn{1}{c}{12} \\
Phase Estimation & 4 & 15 & \multicolumn{1}{c}{21} \\
Quantum Fourier Transform (QFT) & 5 & 4 & \multicolumn{1}{c}{10} \\
Expression Evaluation & 3 & 3 & \multicolumn{1}{c}{7} \\
\bottomrule
\end{tabular}
\end{table}


\textbf{Artificial Benchmarks.} We generated 1000 diverse quantum programs with Qiskit's random circuit generator~\cite{qiskit}. Given the same input, two quantum programs are considered diverse if their outputs differ, which we measure with Jensen Shannon Distance (\jsd). Its range is from 0 to 1, where 0 means the outputs of two programs are exactly the same and 1 means completely different. Note that \jsd has been used in the literature to calculate quantum program diversity based on program outputs~\cite{jsd_approximtion, jsd_approximation2}. For each of the 1000 programs, we calculated the average pairwise \jsd value as its diversity score.
Most benchmarks have an average diversity score of more than 0.5, meaning that on average, each generated circuit, in terms of output, differs more than 50\% from all other circuits.
%
%

We also created three faulty versions of each program in both benchmarks using standard conditional quantum gates such as \textit{cx} and \textit{ccx}. Conditional gates can be configured to activate only on a specific test input, which is convenient for seeding faults in quantum programs in a controllable manner.
To assess the effectiveness of \method, we use the original and the faulty versions of the programs to answer the RQs.


\subsection{Experiment Settings}\label{sec:experimentSetting}
\revision{We use the Qiskit~\cite{qiskit}, Cirq~\cite{cirq}, and pyQuil~\cite{qvm} frameworks for quantum circuit execution. To automatically obtain the program specifications for the benchmarks, we use Qiskit's AER, Cirq's Qsim simulator, and Rigetti's QVM without any noise model to obtain the correct program outputs for \textit{MLP Training}. For noisy quantum circuit execution we used noise models provided by IBM, Google and Rigetti. For IBM, we use Qiskit's AER simulator~\cite{qiskit} integrated with IBM-provided noise models as a noisy backend. Qiskit provides 47 noise models for IBM quantum processors. Out of these 47, we selected 23 based on the criterion that all available noise models must have at least seven qubits for execution since some selected quantum programs require at least seven qubits for their execution. For Google, we use Cirq's Qsim~\cite{cirq} simulator integrated with Google's provided noise models (rainbow and weber) as noisy backend. For Rigetti, we use Rigetti's Quantum Virtual Machine (QVM)~\cite{qvm} with the provided 9 qubit noise model (9q-square) as noisy backend.}

To evaluate the RQs, we split the quantum programs into baseline circuits and CUTs. For the real-world benchmarks, we split them using a 70:30 ratio, i.e., 9 programs split into 3 baseline circuits and 6 CUTs.
Although an 80:20 split ratio is more common, we opt for 70:30 for the real-world benchmarks to have at least three baseline programs to generate enough data for training the baseline models. Note that the 80:20 split ratio cannot be achieved if we want to have three baseline programs. The three baseline programs were then inputted to the \textit{Data Generation} component of \textit{Baseline Trainer}. We split the artificial benchmarks using an 80:20 split ratio, commonly used in ML pipelines~\cite{split,split2, split3}, i.e., 80\% for CUTs and 20\% as baseline circuits for the \textit{Baseline Trainer} component.

For training neural networks, we use the Ktrain~\cite{ktrain} library to construct the training and testing pipeline with mean absolute error as the loss function~\cite{mae, mae2}. The hyper-parameters were automatically selected based on the guideline in~\cite{ktrain}. \revision{After training, \textit{Baseline Trainer} produces 26 baseline MLP models, each of which was fine-tuned in \textit{Baseline Tuner} for each CUT. For fine-tuning the baseline MLP models, we generated data using a maximum of four non-faulty test inputs, depending on the CUT's input space. Although a 7-qubit CUT could have up to 128 test inputs, we intentionally limited it to four, demonstrating that fine-tuning does not require a large number of test inputs. The \textit{Baseline Tuner} results in 156 (6 CUTs, 26 backends) fine-tuned MLP models which were used in the \textit{Noise Filter} component to evaluate the RQs. For the artificial benchmarks, with the 80:20 ratio, we obtain 20800 (800 CUTs, 26 backends) fine-tuned models for \textit{Noise Filter} to evaluate the RQs.}

The experiment was conducted on a machine with a 12th-generation Intel core i9 processor with 20 logical CPUs, 32 GB of RAM, and an Nvidia RTX-3080ti graphics card.

\subsection{Metrics and Statistical Tests}\label{sec:metrics}

To answer RQ1, we used the Hellinger Distance~\cite{hldist}, which is commonly used to compute the similarity of two distributions. In the QC context, it has been used to compare independent execution results of a quantum program on a noisy computer~\cite{Hellinger1,Hellinger2,hellinger3}. For each quantum program, the Hellinger Distance between the program specification ($P$) and the noisy output ($P_N$), or between program specification and the filtered output ($P_F$), is calculated as shown in Eq.~\ref{eq:HL_X}:
%
%
\begin{equation}\label{eq:HL_X}
\hellDist{\ideal}{\X} = \frac{1}{n}\sum_{i=1}^{n}\frac{1}{\sqrt{2}}|\sqrt{P}-\sqrt{X}|
\end{equation}
%
%
where $n$ is the number of inputs of a quantum program, and \X is either $P_N$ or $P_F$. The range of Hellinger Distance is between 0 and 1, where 0 means no difference, and 1 means no similarity. 

For RQ1, we analyze results from two aspects: from the backend aspect, we average the Hellinger distances of all CUTs for each noisy backend, while from the program aspect, we average the Hellinger Distances of all noisy backends for each CUT. 
Moreover, to test statistical significance, based on the guideline~\cite{statistics2,statistics}, we used the Kruskal-Wallis~\cite{kruskal} test for both the program and backend aspects. The Kruskal-Wallis test compares two or more groups in terms of a quantitative variable to test if any group has a clear difference from the other groups. In our case, we use it to check whether \method has significantly different performance in terms of the Hellinger Distance across different backends and programs. Similar performance of most noisy backends indicates that \method can effectively learn and filter noise from different noisy backends. In case of differences in performance, we also use Epsilon Squared effect size and Dunn's test~\cite{Dunn} as recommended posthoc tests for further analyses~\cite{statistics, recomended_effectsize}. For Epsilon Squared effect size, we interpret the values according to the specification given in~\cite{statistics}, where an effect size in the range $[0.01, 0.08)$ is interpreted as \emph{Small}, $[0.08, 0.26)$ is interpreted as \emph{Medium}, and $[0.26, 1]$ is interpreted as \emph{Large}.

In RQ2, we used an existing published test oracle for quantum software testing~\cite{Combinatorial, search}. We compared the test assessment results for the original and faulty programs with and without our approach across all noisy backends. We also used the Mann-Whitney~\cite{mann} statistical test and Vargha Delaney \Atwelve~\cite{A12} effect size as recommended in~\cite{recomended_effectsize,statistics2} for studying the statistical significance of the results. The Mann-Whitney test is used to compare differences between two independent groups. In our case, we use it to compare the ideal test case assessment with \method integrated test case assessment. To evaluate the magnitude of differences between these two groups, \Atwelve~\cite{A12} is interpreted according to~\cite{statistics}, where an effect size in the range
$(0.34, 0.44]$ and $[0.56, 0.64)$ is interpreted as \emph{Small}, $(0.29, 0.34]$ and $[0.64, 0.71)$ is interpreted as \emph{Medium}, and $[0, 0.29]$ and $[0.71, 1]$ is interpreted as \emph{Large}.

For RQ2, an existing test oracle for quantum software testing is used to perform test case assessments on all possible test inputs for the CUTs. Each CUT is assigned a score at the end of the test case assessments: \textit{Score\%}, defined as the average percentage (across all noisy backends) of test inputs for which the CUT failed to pass the test oracle. Test case assessment results for each input are classified as True Positive (TP), False Negative (FN), False Positive (FP), or True Negative (TN), as follows: 
\textbf{TP} - The program specification says faulty, and test case assessment also says faulty; 
\textbf{FP} - Test case assessment says faulty, but the program specification says not faulty; 
\textbf{FN} - test case assessment says not faulty, but the program specification says faulty; and 
\textbf{TN} - The program specification says not faulty, and test case assessment also says not faulty.
We calculate F-measure (F1 score) (combining the statistics of precision and recall)~\cite{commonf1, commonf2, commonf3} for the program specification of all the test inputs and compare them with both noisy test case assessment results and filtered test case assessment results, with the formula below:
\begin{equation}\label{eq:F1}
F1 = 2 * \frac{(\mathit{Precision} \times \mathit{Recall})}{(\mathit{Precision} + \mathit{Recall})},
\end{equation}
where \textit{Precision} is given by TP/(TP+FP) and \textit{Recall} is given by TP/(TP+FN). The F1 score ranges from 0 to 1, where 0 means all the test case assessments are inconsistent, and 1 means all are consistent.

\section{Results and Discussion}\label{sec:results}

\subsection{RQ1 -- Noise effect reduction} \label{subsec:RQ1}
RQ1 assesses \method regarding the accuracy and quality of the trained neural network in reducing the noise effect from a noisy backend. Accuracy is measured with Hellinger Distance (see Sect.~\ref{sec:metrics}).

\subsubsection{Results from the backend aspect}

Table~\ref{table:RQ1_distance_backend} shows the average results of the Hellinger Distances from the backend aspect for the real-world and artificial benchmarks.
\begin{table}[!tb]
\caption{RQ1 (the backend aspect) -- Results of the average Hellinger Distance for each backend across all CUTs. Column $\boldsymbol{\hellDist{i}{n}}$ (or $\boldsymbol{\hellDist{i}{f}}$) shows the average Hellinger Distance between the ideal outputs specified in the program specification and noisy (or filtered) program outputs for each noisy backend across all CUTs. Column $\boldsymbol{Improved \%}$ is the percentage change between $\boldsymbol{\hellDist{i}{n}}$ and $\boldsymbol{\hellDist{i}{f}}$ calculated as $\frac{(\boldsymbol{\hellDist{i}{n}}-\boldsymbol{\hellDist{i}{f}})}{\boldsymbol{\hellDist{i}{n}}}*100$.}
\label{table:RQ1_distance_backend}
\centering
\setlength{\tabcolsep}{2pt}
\begin{tabular}{c|ccc|ccc}
\toprule
\multirow{2}{*}{\textbf{backend}} & \multicolumn{3}{c|}{\textbf{Real-world Benchmarks}} & \multicolumn{3}{c}{\textbf{Artificial Benchmarks}} \\ \cline{2-7} 
& $\boldsymbol{\hellDist{i}{n}}$ & $\boldsymbol{\hellDist{i}{f}}$ & \multicolumn{1}{c|}{$\mathit{Improved}\%$} & $\boldsymbol{\hellDist{i}{n}}$ & $\boldsymbol{\hellDist{i}{f}}$ & $\mathit{Improved}\%$ \\
\midrule
Almaden & 0.47 & 0.08 & \textbf{83.85} & 0.22 & 0.04 & \textbf{81.01} \\
Boeblingen & 0.45 & 0.06 & \textbf{86.82} & 0.21 & 0.04 & \textbf{81.70} \\
Brooklyn & 0.40 & 0.05 & \textbf{86.80} & 0.17 & 0.03 & \textbf{81.15} \\
Cairo & 0.33 & 0.04 & \textbf{88.91} & 0.15 & 0.03 & \textbf{81.29} \\
Cambridge & 0.54 & 0.25 & 53.33 & 0.31 & 0.08 & 73.59 \\
CambridgeV2 & 0.54 & 0.26 & 52.02 & 0.31 & 0.08 & 73.71 \\
Casablanca & 0.42 & 0.05 & \textbf{87.19} & 0.17 & 0.03 & \textbf{82.55} \\
Guadalupe & 0.39 & 0.06 & \textbf{84.08} & 0.16 & 0.03 & \textbf{80.90} \\
Hanoi & 0.29 & 0.03 & \textbf{89.94} & 0.14 & 0.03 & \textbf{80.48} \\
Jakarta & 0.41 & 0.06 & \textbf{86.18} & 0.17 & 0.03 & \textbf{82.42} \\
Johannesburg & 0.49 & 0.10 & 78.80 & 0.26 & 0.04 & \textbf{84.77} \\
Kolkata & 0.32 & 0.03 & \textbf{91.14} & 0.13 & 0.03 & 77.99 \\
Lagos & 0.32 & 0.04 & \textbf{88.96} & 0.12 & 0.03 & 78.23 \\
Manhattan & 0.44 & 0.06 & \textbf{87.38} & 0.27 & 0.10 & 62.00 \\
Montreal & 0.34 & 0.05 & \textbf{84.84} & 0.15 & 0.03 & \textbf{80.02} \\
Mumbai & 0.35 & 0.04 & \textbf{88.78} & 0.17 & 0.03 & \textbf{83.38} \\
Nairobi & 0.38 & 0.05 & \textbf{87.99} & 0.17 & 0.03 & \textbf{82.45} \\
Paris & 0.37 & 0.05 & \textbf{86.72} & 0.17 & 0.03 & \textbf{80.70} \\
Rochester & 0.59 & 0.38 & 35.14 & 0.40 & 0.13 & 67.11 \\
Singapore & 0.45 & 0.07 & \textbf{85.20} & 0.24 & 0.04 & \textbf{82.56} \\
Sydney & 0.36 & 0.05 & \textbf{86.92} & 0.18 & 0.03 & \textbf{82.30} \\
Toronto & 0.57 & 0.42 & 26.49 & 0.30 & 0.09 & 69.41 \\
Washington & 0.34 & 0.04 & \textbf{89.23} & 0.16 & 0.03 & \textbf{81.77} \\
\revision{9q-square} & \revision{0.66} & \revision{0.39} & \revision{40.40} & \revision{0.54} & \revision{0.28} & \revision{48.08} \\
\revision{Rainbow} & \revision{0.68} & \revision{0.30} & \revision{56.24} & \revision{0.58} & \revision{0.31} & \revision{46.78} \\
\revision{Weber} & \revision{0.70} & \revision{0.49} & \revision{29.87} & \revision{0.59} & \revision{0.31} & \revision{46.96} \\
\bottomrule
\end{tabular}
\end{table}
Column $\boldsymbol{\hellDist{i}{n}}$ shows the average difference between the results of the ideal program outputs specified in the program specification and noisy program outputs for each noisy backend across all CUTs; $\boldsymbol{\hellDist{i}{f}}$ shows the difference between results of the ideal program outputs in the program specification and filtered program outputs for each noisy backend across all CUTs, and $\boldsymbol{Improved \%}$ shows the percentage improvement from $\boldsymbol{\hellDist{i}{n}}$ to $\boldsymbol{\hellDist{i}{f}}$. \revision{From the table, we can see that for the real-world benchmarks, on 18 out of 26 backends, \method achieved an improvement of over $80\%$, whereas on backend \textit{Cambridge}, \textit{CambridgeV2\footnote{Qiskit provides two Cambridge backends: Cambridge, and CambridgeV2. These two backends are essentially the same regarding hardware configuration, with the only difference being the measurement basis.}}, \textit{Rochester}, \textit{9q-square}, \textit{Rainbow}, \textit{Weber}, and \textit{Toronto}, \method achieved the least improvement. 
For the artificial benchmarks, on 16 backends, \method achieved an improvement of more than $80\%$ whereas, on backends \textit{Cambridge}, \textit{CambridgeV2}, \textit{Rochester}, \textit{9q-square}, \textit{Rainbow}, \textit{Weber}, \textit{Toronto} and \textit{Manhattan}, \method performed the worst. For both the real-world and artificial benchmarks, on backends \textit{Cambridge}, \textit{CambridgeV2}, \textit{Rochester}, \textit{9q-square}, \textit{Rainbow}, \textit{Weber}, and \textit{Toronto}, \method achieved the least improvement, indicating that these backends have more considerable noise effect that is hard to be filtered out by \method than the others for the selected CUTs. Considering all backends, the overall average improvement for real-world programs is 74.7\%, and for artificial ones, it is 75.1\%, which shows that \method effectively filtered out the noise effect from the program outputs produced by most noisy backends.}

To test whether there is a significant difference in the performance of \method across backends, we present the results of the Kruskal-Wallis test and the Epsilon-Squared effect size. \revision{For the real-world benchmarks, from the backend aspect, the p-value is close to \textbf{0}, less than $\boldsymbol{0.05}$ with a large range effect size $\boldsymbol{0.32}$, indicating that there is a significant difference among the backends considering the Hellinger Distance. To evaluate which backend pairs have significant differences, we performed the Dunn's test for post-hoc analysis. Results show that the \textit{Weber} backend from Google has significant differences with at least four other backends. 
For the artificial benchmarks, the p-value of the Kruskal-Wallis test is also close to $\boldsymbol{0}$, less than the $\boldsymbol{0.05}$ with a large range effect size of $\boldsymbol{0.44}$ (see Sect.~\ref{sec:metrics}).
This shows that in terms of Hellinger Distance, significant differences exist among the noisy backends for the artificial benchmarks. Dunn's test results show that each of the seven noisy backends (i.e., Rochester, Cambridge, CambridgeV2, Toronto, 9q-Square, Rainbow and Weber) had significant differences with all other backends, which can also be seen in Table~\ref{table:RQ1_distance_backend} where these backends have the least improvement. Due to space limitations, the complete results are provided in the online repository~\cite{sourcecode}}.

\subsubsection{Results from the program aspect} 
Table~\ref{table:RQ1_distance_program} shows the average results of Hellinger Distance difference from the program aspect for the real-world benchmarks.
\begin{table}[!tb]
\caption{RQ1 (the program aspect) -- Results of the average Hellinger Distance for each CUT across all noisy backends for the real-world benchmarks. Column $\boldsymbol{\hellDist{i}{n}}$ (or $\boldsymbol{\hellDist{i}{f}}$) denotes the average Hellinger Distance between the ideal outputs specified in the program specification and noisy (or filtered) outputs for each CUT across all noisy backends; Column $\boldsymbol{Improved \%}$ is the percentage change between the $\boldsymbol{\hellDist{i}{n}}$ and $\boldsymbol{\hellDist{i}{f}}$, calculated as $\frac{(\boldsymbol{\hellDist{i}{n}}-\boldsymbol{\hellDist{i}{f}})}{\boldsymbol{\hellDist{i}{n}}}*100$.}
\label{table:RQ1_distance_program}
\centering
\begin{tabular}{c|ccc}
\toprule
\textbf{\textit{CUT}} & $\boldsymbol{\hellDist{i}{n}}$ & $\boldsymbol{\hellDist{i}{f}}$ & $\boldsymbol{Improved \%}$ \\
\midrule
\multicolumn{1}{c|}{GHZ} & \multicolumn{1}{c}{\revision{0.299}} & \multicolumn{1}{c}{\revision{0.002}} & \textbf{\revision{99.2}} \\
\multicolumn{1}{c|}{Simon} & \multicolumn{1}{c}{\revision{0.152}} & \multicolumn{1}{c}{\revision{0.055}} & \textbf{\revision{63.9}} \\
\multicolumn{1}{c|}{QFT} & \multicolumn{1}{c}{\revision{0.664}} & \multicolumn{1}{c}{\revision{0.142}} & \textbf{\revision{78.5}} \\
\multicolumn{1}{c|}{Addition} & \multicolumn{1}{c}{\revision{0.573}} & \multicolumn{1}{c}{\revision{0.138}} & \textbf{\revision{75.8}} \\
\multicolumn{1}{c|}{Binary Similarity} & \multicolumn{1}{c}{\revision{0.672}} & \multicolumn{1}{c}{\revision{0.219}} & \textbf{\revision{67.4}} \\
\multicolumn{1}{c|}{Phase Estimation} & \multicolumn{1}{c}{\revision{0.313}} & \multicolumn{1}{c}{\revision{0.245}} & \revision{21.6} \\
\bottomrule
\end{tabular}
\end{table}
\revision{From the table, we can observe that, on average, for 5 out of the 6 CUTs, \method achieved more than $60\%$ improvement. For program \textit{Phase Estimation}, \method, however, only achieved an improvement of $21.6\%$. 
For the 800 artificial benchmarks, due to space limitation, detailed results are provided in the online repository~\cite{sourcecode}, and we only summarize key findings as follows. For 155 out of the 800 artificial benchmarks, \method achieved an average improvement of more than $\boldsymbol{80\%}$, 281 programs have more than $\boldsymbol{70\%}$, 203 programs have more than $\boldsymbol{50\%}$, and 142 programs have more than $\boldsymbol{10\%}$ average improvement; for the other 19 programs, \method achieved a very minor or no improvement.}

\revision{A possible explanation for the exceptions of the \textit{Phase Estimation} program and the 19 artificial programs is that they all consist of more phase gates than other types. This indicates that the noise effect of different noisy backends on programs that mostly consist of phase gates is challenging to distinguish and requires further analysis. One possible reason could be the train-test split of the baseline circuits and CUTs for the \textit{Baseline Training} component. We selected baseline circuits randomly with a specific split ratio (see Sect.~\ref{sec:experimentSetting}). After observing the results for \textit{Phase Estimation} and the 19 artificial programs, we checked the baseline circuits used for training in \textit{Baseline Trainer} and noticed that the selected baseline circuits have fewer phase gates than other types of quantum gates. This could lead to a bias in training, as the training dataset captures the accumulating effect of noise and not individual gate noise. As a result, the imbalance between the number of phase gates and other types of quantum gates could affect the performance of \method on programs with many phase gates. In future studies, we will also focus on better circuits for MLP training.}

\begin{figure}[!tb]
\includegraphics[width=6cm]{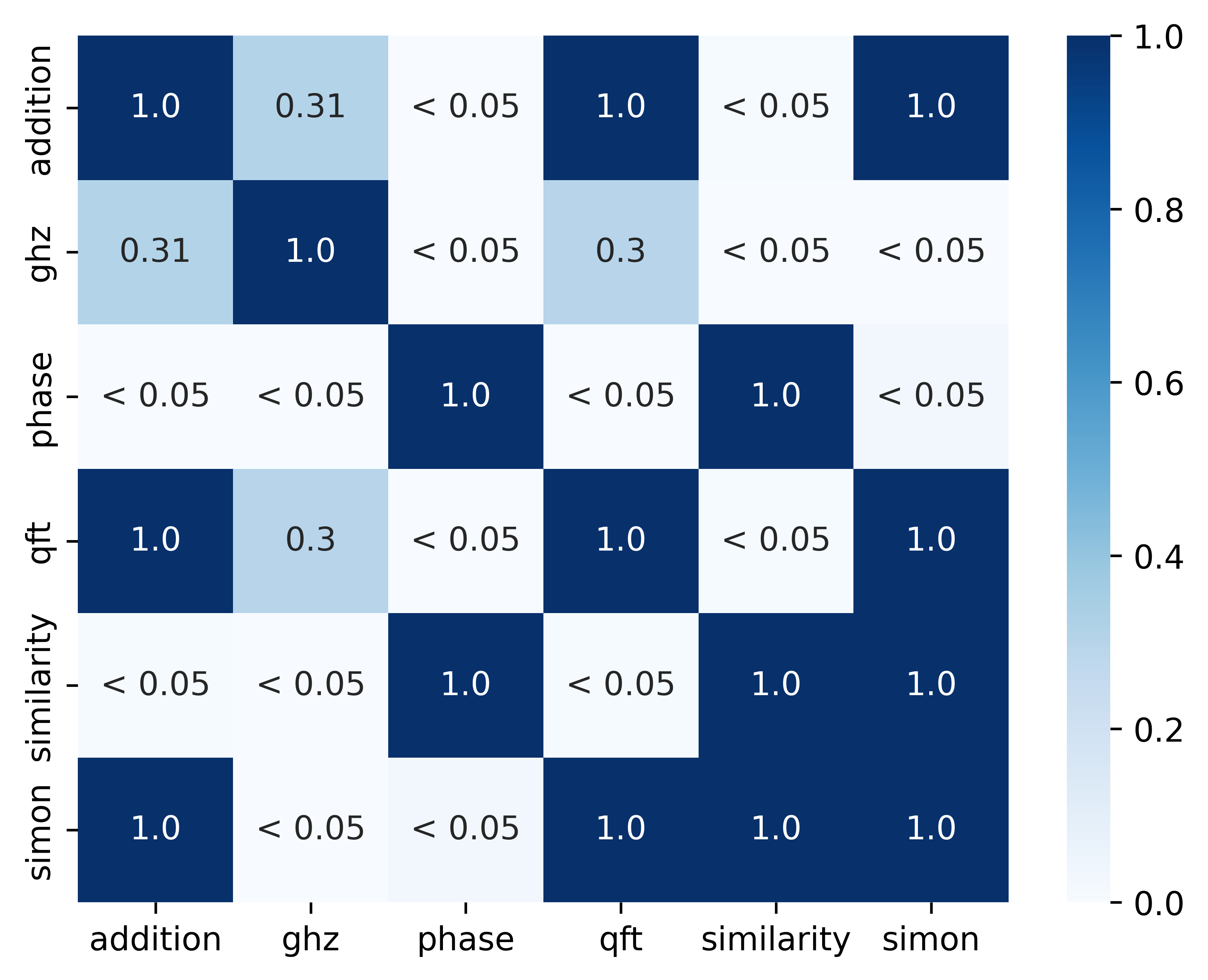}
\centering
\caption{RQ1 (the program aspect) -- Dunn's test results for the real-world benchmarks. The darker blue coloring shows that the magnitude of noise effect reduction in Hellinger Distance is more similar between a pair of programs.}
\label{fig:posthoc}
\end{figure}

In terms of statistical test results of the Kruskal-Wallis test and Epsilon-Squared effect size, for the real-world benchmarks, \revision{the p-value is $\boldsymbol{1.27e^{-13}}$ which is less than $\boldsymbol{0.05}$ with an effect size of $\boldsymbol{0.44}$, indicating that there is at least one program for which \method performed significantly different than for the others across all noisy backends. We further checked the results of Dunn's test (Fig.~\ref{fig:posthoc}) for the real-world benchmarks.}
From Fig.~\ref{fig:posthoc}, we can see that the \textit{Phase Estimation} program clearly differs from the other programs, which is consistent with what we observed in Table~\ref{table:RQ1_distance_program}. We also see that the \textit{Similarity} program is also similar to \textit{Simon} and \textit{Phase Estimation}, whereas \textit{Addition}, \textit{GHZ}, and \textit{QFT} are similar among each other. One possible reason could be that the circuit structure (in terms of phase gates) of the \textit{Similarity} program is more similar to \textit{Phase Estimation} and \textit{Simon} than to other programs. Similarly, programs \textit{Addition}, \textit{GHZ}, and \textit{QFT} (that have no phase gates) are similar to each other. After \textit{Phase Estimation}, the \textit{Similarity} program has the highest value for $\hellDist{i}{f}$ because it has more phase gates than the other programs, which supports the hypothesis that noise introduced by phase gates is more difficult to distinguish than for other quantum gates.

Results of the Kruskal-Wallis test and Epsilon-Squared effect size for the artificial benchmarks show that the p-value is close to $\boldsymbol{0}$ (less than $\boldsymbol{0.05}$) with an effect size of $\boldsymbol{0.40}$, indicating a significant difference regarding \method's performance among the artificial benchmarks. Dunn's test results show that 19 out of the 800 artificial benchmarks (consisting of phase gates) have significant differences among themselves and with the others. Detailed results are available in the online repository~\cite{sourcecode}.
This observation asks if \method can generalize noise learning among similar programs for all noisy backends. To answer this question, we identify groups of similar artificial benchmarks based on their output diversity scores (\jsd) (see Sect.~\ref{sec:benchmarks}) and then quantify the average pairwise difference for \hellDist{i}{f} (Table~\ref{table:RQ1_distance_program}) for programs in each group across all noisy backends. To identify program groups, we rounded the diversity score to one decimal place and identified four scores ranging from 0.5 to 0.8. Table~\ref{tab:RQ3} presents the results.
\begin{table}[!tb]
\caption{RQ2 -- Average pairwise differences in Hellinger Distance (column \textit{Avg. Pairwise HL}) for the programs with a similar diversity score. The first column shows the average diversity score (in \jsd) groups, and the second column is the number of CUTs belonging to each group.}
\label{tab:RQ3}
\centering
\begin{tabular}{ccc}
\toprule
\textbf{\textit{Avg. Diversity Score Group}} & \textbf{\textit{\# Programs}} & \textbf{\textit{Avg. Pairwise HL}} \\
\midrule
0.5 & 46 & 0.025 \\
0.6 & 485 & 0.031 \\
0.7 & 236 & 0.038 \\
0.8 & 16 & 0.028 \\
\bottomrule
\end{tabular}
\end{table}
As shown in Table~\ref{tab:RQ3}, 485 out of the 800 CUTs have an average diversity score of 0.6, whereas only 16 CUTs have a diversity score of 0.8. Like Table~\ref{table:RQ1_distance_program}, for each program in each average diversity score group, its \hellDist{i}{f} values across all noisy backends are averaged.
For \method to have a similar magnitude of noise effect reduction for a particular group of programs, the average pairwise difference in \hellDist{i}{f} should be as close to zero as possible. Table~\ref{tab:RQ3} shows that the average pairwise difference in \hellDist{i}{f} for all groups is greater than 0.02 but less than 0.04, with the maximum value being 0.038. 
This shows that for any pair of programs with similar output distributions, \method generalizes the amount of noise effect it filters from the programs' output.

\subsubsection{\revision{Variance of ML models}}
\begin{figure}[!tb]
\includegraphics[width=8cm]{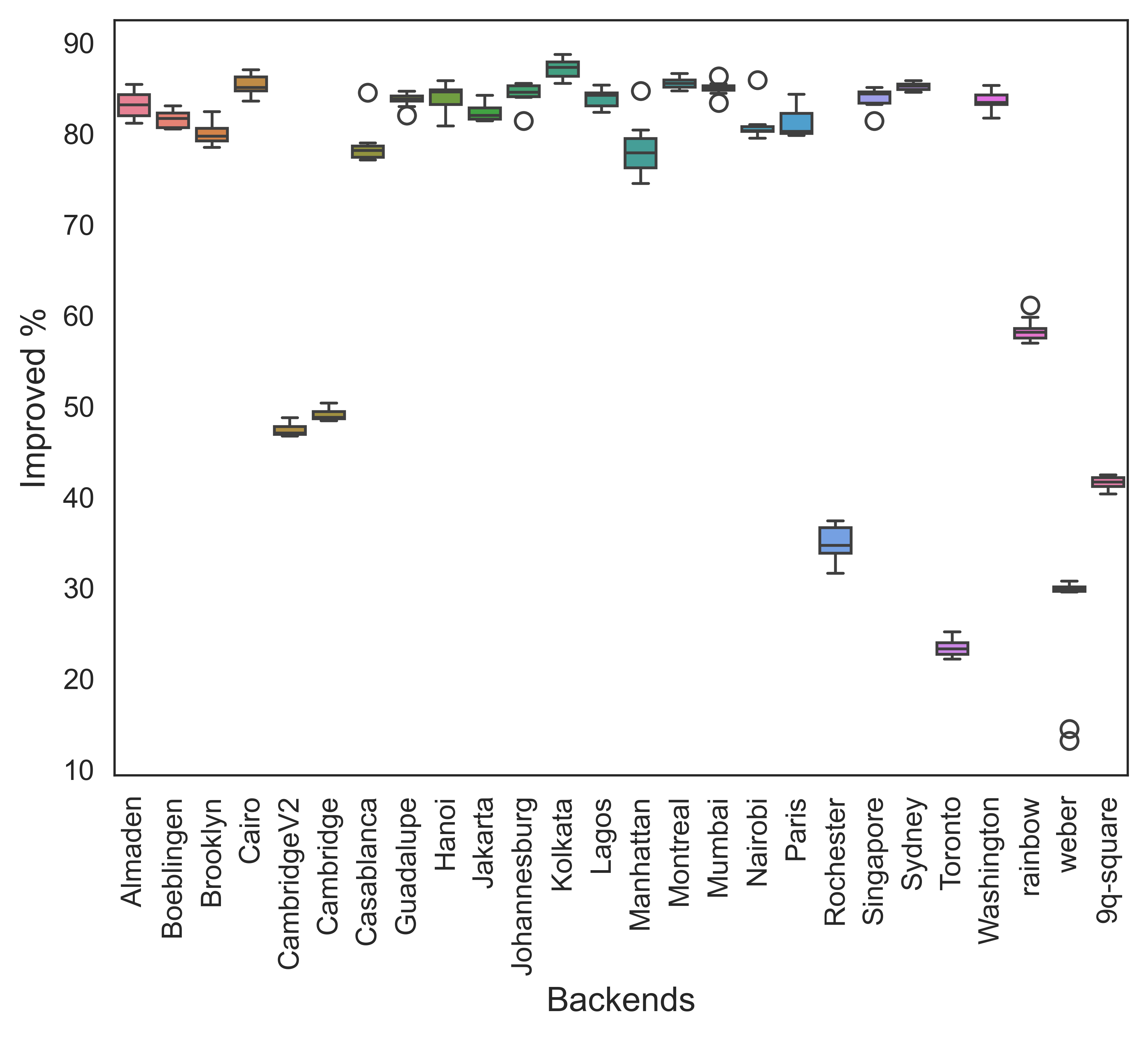}
\centering
\caption{\revision{RQ1 -- Result of \textit{Improved\%}~(see Table~\ref{table:RQ1_distance_backend}) on the real-world benchmark for 10 runs. Each box plot shows the distribution of percentage improvement achieved by \method for each backend for 10 runs.}}
\label{fig:RQ1dsit}
\end{figure}

\revision{ML models are inherently probabilistic, leading to multiple predicted filtered outputs for multiple noisy program outputs. To assess the extent of variation in ML model predictions with changes in noisy program output, we computed the metrics detailed in Table~\ref{table:RQ1_distance_backend} over ten runs, only for the real-world benchmarks because a huge amount of time would be required if computing for all benchmarks. Results are reported in Fig.~\ref{fig:RQ1dsit}.}
\revision{Fig.~\ref{fig:RQ1dsit} demonstrates a consistent alignment of results with the average outcomes presented in Table~\ref{table:RQ1_distance_backend} for each backend. The observed variance, in general, is minimal for most backends, suggesting that the ML models consistently predicted similar results for different noisy program outcomes across various runs.}
\begin{tcolorbox}[colback=blue!5!white,colframe=white,
breakable
]
\textbf{Answer to RQ1:} \method can effectively reduce the noise effect on the outputs of a quantum program running on most noisy backends by more than $80\%$. With a pairwise difference of less than 0.04 for a group of similar quantum programs, \method also generalizes the amount of noise effect it filters out from outputs of quantum programs.
\end{tcolorbox}
\subsection{RQ2 -- Test case assessment improvement} \label{subsec:RQ2}
To answer RQ2, we integrated \method with an existing test oracle defined and used in two quantum software testing methods~\cite{Combinatorial, search}, based on which the test case assessment was performed for test execution results on the original programs and their three faulty versions across all noisy backends. 

\subsubsection{Categorizing noise effect migrations in test case assessments}\label{subsec:migratio}
\begin{figure}[!tb]
\includegraphics[width=8.5cm]{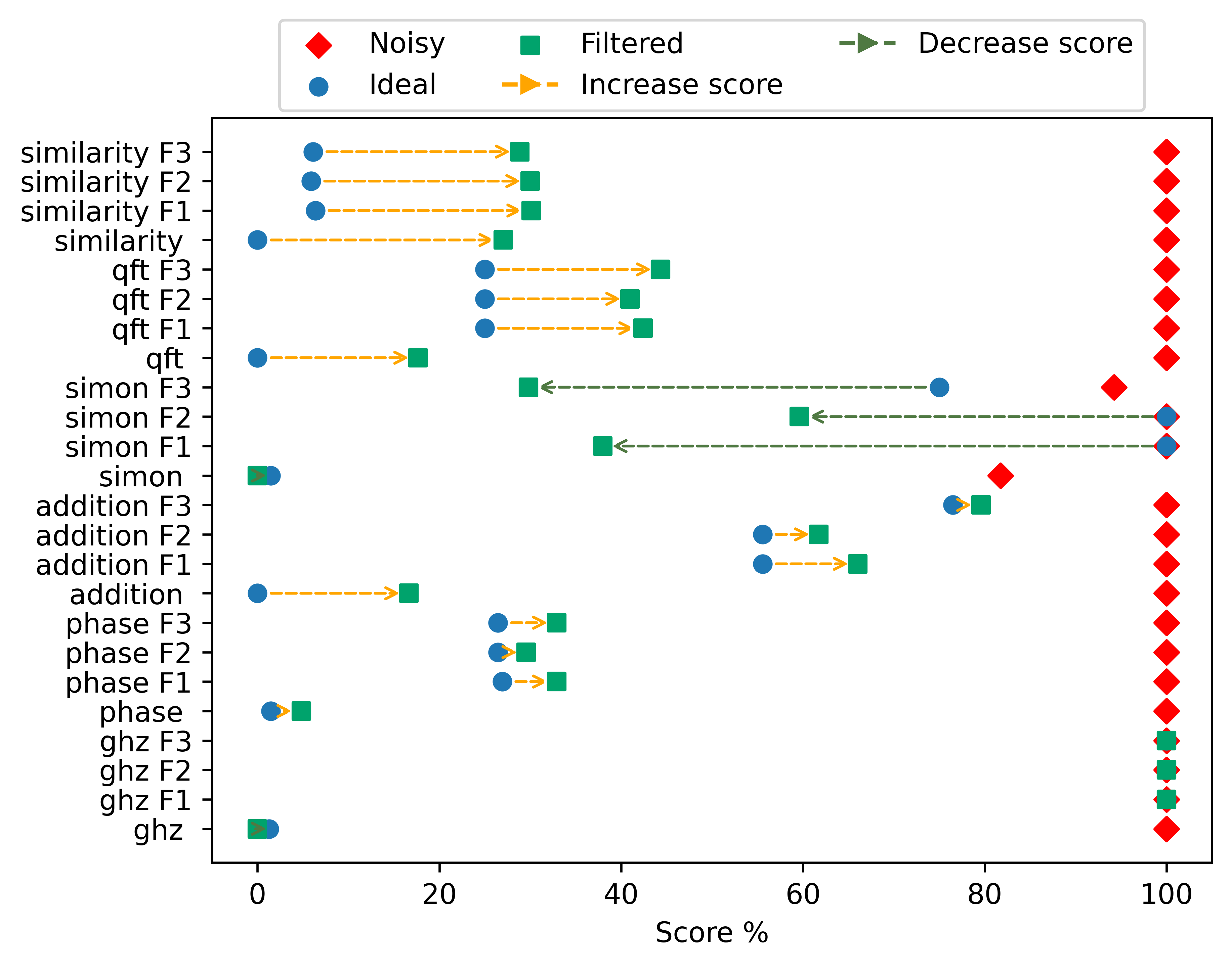}
\centering
\caption{RQ2 -- Results of the test case assessment for the original and its faulty programs of the real-world benchmarks across all noisy backends on average. The x-axis (\textit{Score \%}) is the average percentage of inputs on which a program failed the test case assessment across all backends; the y-axis labels the original programs (e.g., \textit{qft}) and its three faulty ones (e.g., \textit{qft F1}, \textit{qft F2} and \textit{qft F3}).}
\label{fig:RQ2B1}
\end{figure}

Fig.~\ref{fig:RQ2B1} shows the average results of the test case assessments for all noisy backends on the original and faulty programs (with suffixes of F1, F2, F3) of the real-world benchmarks.
\revision{
From the Figure~\ref{fig:RQ2B1}, we see that four out of six original original programs (four blue dots located at the 0\% position on the x-axis) exhibit no fault. The remaining two original programs (\textit{Phase and Simon}) are closer to 0\% but not exactly 0\% due to false positives for some inputs. Assessment results for the noisy backends without \method (denoted with red diamonds) indicate that most programs failed on 100\% of the inputs. The green squares denote the test case assessment results after applying \method, and the dashed arrows visualize the difference between the ideal outputs and the results after applying \method. We can clearly see that \method allowed test case assessment with the noisy backends much closer to the program specifications, as the yellow dashed lines connecting results of the program specifications (ideal) and noisy backends with \method applied are very short.} 
Fig.~\ref{fig:RQ2B1} shows three cases: \revision{\textbf{\textit{Case-1}--no difference} in the scores of the ideal outputs and results from the noisy backend and filtered with \method, illustrated as overlapped blue dots and green squares, i.e., \textit{ghz F1}, \textit{ghz F2} and \textit{ghz F3} in the figure;} \textbf{\textit{Case-2}--decrease} in the scores regarding the noise effect migration from the ideal outputs to the noisy backend, indicating fewer inputs failed and more false negatives in the test case assessment results with \method on the noisy backend (e.g., the dashed green arrow on \textit{simon F3}); \textbf{\textit{Case-3}--increase} in the scores regarding the noise effect migration from the ideal outputs to the noisy backend (e.g., the dashed yellow arrow on \textit{addition}), telling that more inputs failed and more false positives observed in filtered test case assessment results.

Test inputs can have different results on different noisy backends; therefore, we use the Mann-Whitney statistical test and Vargha Delaney A12 effect size~\cite{A12} to compare the scores achieved by comparing the ideal outputs in the program specification with those from the noisy backends and filtered with \method for all noisy backends. 
From Table~\ref{tab:RQ2B1B2}, we can observe that four out of the six real-world benchmarks fall into Case-3, among which only one exhibits a large effect size on the significant difference between the ideal outputs in the program specifications and results from the noisy backends filtered with \method across all noisy backends.
\begin{table}[!tb]
\caption{RQ2 -- Summary of the number of quantum programs falling into the three noise effect migration cases: Case-1--No difference; Case-2--Decrease from the ideal to filtered; Case-3 - Increase from the ideal to filtered. Row 
\textit{Effect (Large)} shows the number of programs that have a large effect size (Vargha-Delaney A12).}
\label{tab:RQ2B1B2}
\centering
\setlength{\tabcolsep}{4pt}
\begin{tabular}{c|cccc}
\toprule
\multirow{2}{*}{} & \multicolumn{2}{c}{\textbf{Real-world Benchmarks}} & \multicolumn{2}{c}{\textbf{Artificial Benchmarks}} \\ \cline{2-5} 
& \multicolumn{1}{c|}{$\boldsymbol{Original}$} & \multicolumn{1}{c|}{$\boldsymbol{Faulty}$} & \multicolumn{1}{c|}{$\boldsymbol{Original}$} & \multicolumn{1}{c}{$\boldsymbol{Faulty}$} \\
\midrule
Case-1 & \revision{0} & \multicolumn{1}{c|}{\revision{3}} & \revision{93} & \multicolumn{1}{c}{\revision{79}} \\
Case-2 & \revision{2} & \multicolumn{1}{c|}{\revision{3}} & \revision{0} & \multicolumn{1}{c}{\revision{581}} \\
Case-3 & \revision{4} & \multicolumn{1}{c|}{\revision{12}} & \revision{707} & \multicolumn{1}{c}{\revision{1740}} \\ \hline
Effect (Large) & 1 & \multicolumn{1}{c|}{6} & \revision{11} & \multicolumn{1}{c}{\revision{505}} \\
\bottomrule
\end{tabular}
\end{table}
For the 18 faulty programs, only 6 have a large effect size, for which the ideal outputs are significantly better than that from the noisy backend and filtered by \method across all noisy backends.
%
\revision{For the artificial benchmarks, only eleven of the 800 original programs show a large effect size on the significance when comparing the ideal outputs and results from the noisy backends filtered with \method across all noisy backends, and 21\% (505 out of 2400) of the faulty programs exhibit significant differences.}

\subsubsection{Calculating F1-score of \method in test case assessment}
In the real-world and artificial benchmarks, we observed many programs changed in \textit{Score\%} (see Sect.~\ref{sec:metrics}), though only a small portion shows a large effect size on the significance (Table~\ref{tab:RQ2B1B2}). Such changes occurred due to false positives (Case-3) or false negatives (Case-2) produced by \method. To evaluate the quality of \method in reducing false positives or false negatives, we calculate the F-measure (F1-score). A typical test case assessment is similar to binary classification in that each assessment results in either a fault or not. We compare the F1-score (see Eq.~\ref{eq:F1}) for the test case assessment with \method and without \method for all programs across all noisy backends. Results are summarized in Table~\ref{tab:RQ2F}.
%
\begin{table}[!tb]
\caption{RQ2 -- F1-score, Precision, and Recall for all test inputs across all noisy backends. A higher F1-score means a lower chance of having false positives or negatives, a higher precision means fewer false positives, and a higher recall means fewer false negatives. Columns \textit{w/o \method} and \textit{with \method} show the results for all test inputs across all noisy backends without \method and \textit{with \method}, respectively.}
\label{tab:RQ2F}
\centering
\setlength{\tabcolsep}{2pt}
\begin{tabular}{c|cccc}
\toprule
\multirow{2}{*}{\textbf{}} & \multicolumn{2}{c}{\textbf{Real-world Benchmarks}} & \multicolumn{2}{c}{\textbf{Artificial Benchmarks}} \\ \cline{2-5} 
& \textbf{w/o \method} & \multicolumn{1}{c|}{\textbf{with \method}} & \textbf{w/o \method} & \textbf{with \method} \\
\midrule
\textbf{F1-score} & \revision{0.02} & \multicolumn{1}{c|}{\revision{\textbf{0.86}}} & \revision{0.07} & \revision{\textbf{0.86}} \\ \cline{1-1}
\textbf{\revision{Precision}} & \revision{1.0} & \multicolumn{1}{c|}{\revision{\textbf{0.99}}} & \revision{1.0} & \textbf{\revision{0.93}} \\ \cline{1-1}
\textbf{\revision{Recall}} & \revision{0.01} & \multicolumn{1}{c|}{\revision{\textbf{0.75}}} & \revision{0.03} & \textbf{\revision{0.79}} \\ \cline{1-1}
\textbf{False Positives} & \revision{0} & \multicolumn{1}{c|}{\revision{\textbf{22}}} & \revision{0} & \textbf{\revision{32,237}} \\ \cline{1-1}
\textbf{False Negatives} & \revision{11,689} & \multicolumn{1}{c|}{\revision{2,946}} & \revision{553,965} & \revision{117,485} \\ 
\bottomrule
\end{tabular}
\end{table}

\revision{In Table~\ref{tab:RQ2F}, a substantial improvement in the F1 score is evident after the integration of \method. For the real-world benchmarks, the F1 score increased from 2\% to 86\%, and for the artificial benchmarks, it improved from 7\% to 0.86\%. Generally, a high number of false positives and false negatives in test case assessments indicates that the assessment is unreliable. In Table~\ref{tab:RQ2F}, we can observe a notable reduction in the number of false negatives with \method (from 11,689 to 2,946 for real-world benchmarks and from 553,965 to 117,485 for artificial benchmarks). However, it is important to note that \method also introduces false positives, indicating instances where it was unable to filter out noise from some test inputs. Considering precision, in both benchmarks, we observe that without \method, precision is 1.0, indicating no false positives occurred. This is because all test inputs failed, including the ones meant to fail. On the other hand, recall for real-world and artificial benchmarks without \method is 0.01 and 0.03 respectively which is very low, indicating a high number of false negatives. With \method, we can see a significant improvement in recall for both benchmarks, while precision is slightly decreased, indicating that for some test inputs, \method was not able to reduce noise. Overall, the F1 score of \method is close to the ideal F1 score of 1.0 and demonstrates significantly fewer false negatives. This suggests that \method can be a valuable tool for enhancing the test assessment of existing methods for noisy backends.}

\begin{tcolorbox}[colback=blue!5!white,colframe=white,
breakable
]
\textbf{Answer to RQ2:} \method can be integrated effectively with existing test methods for testing quantum programs on a noisy backend, and \method can reduce the gap between the program specification and test case assessments on program outputs from the noisy backend with precision, recall, and F1-score of 99\%, 75\%, and 85\%, respectively, for real-world benchmarks. Similarly, for artificial benchmarks, it demonstrates a precision, recall, and F1-score of 93\%, 79\%, and 85\%, respectively.
\end{tcolorbox}

\subsection{Threats to validity}\label{sec:threats}
\textbf{External validity:} We selected a small set of real-world quantum programs for evaluating \method; therefore, it threatens the generalizability of the results. Though we understand that more programs help improve generalizability, we only found a handful of real-world quantum programs that can be used for our evaluation. To mitigate this threat, we generated 1000 artificial quantum programs by following a common practice from~\cite{Qdiff, morphq}. We also selected quantum programs/circuits that operate on a small number of qubits, which may also affect the generalizability. Indeed, the effect of noise depends on the number of qubits, i.e., a higher number of qubits will increase the chances of errors due to noise. However, we cannot choose quantum programs with a large number of qubits since doing so will restrict the selection of noisy backends to use and make the programs requiring a large number of qubits impossible to execute on all noisy backends. Out of 47 available noisy backends from IBM, only 23 have a qubit count of at least seven, which we chose in our experiments. The next threat is related to improper optimization of hyperparameters of ML algorithms. ML models can easily overfit the training data, affecting the model's generalizability. To reduce this threat, we followed the guidelines of~\cite{ktrain, learningrate} to construct our machine-learning pipeline for auto-tuning the parameters and monitoring the training loop to avoid over-fitting.

\textbf{Construct Validity:} To reduce the threats concerning the evaluation metrics, we used Hellinger Distance (see Sect.~\ref{sec:metrics}) to measure the diversity of outputs of quantum programs, which has been widely used to compare the outputs of quantum programs~\cite{Hellinger1, Hellinger2, hellinger3}. We also used F1-score, precision and recall to measure the quality of \method in noise effect reduction, standard quality metrics used to evaluate binary classification problems~\cite{commonf1,commonf2, commonf3}. For statistical tests, we followed the guidelines provided in~\cite{statistics2, statistics, recomended_effectsize}. 

\textbf{Conclusion Validity:}
Not applying \method on real NISQ computers is a threat to the conclusion validity of the evaluation. Available NISQ computers offer limited public access, making it infeasible to evaluate \method on them. To mitigate this threat, \revision{we used noise models provided by IBM, Google, and Rigetti for their real quantum computers. Each noise model closely approximates the behavior of a quantum processor and is updated frequently.}
\revision{To demonstrate the effectiveness of \method, we incorporated 26 noise models matching our inclusion criteria at the time of the experiment. With \method, a tester can effectively reduce noise effect from program outputs (i.e., more than 80\%, see Sect.~\ref{subsec:RQ1}), thereby enabling the tester to conclude with a certain degree of confidence on whether a failure was observed due to a fault in a program or hardware noise (i.e., improved test assessment F1-score by, at least, 85\%, see Sect.~\ref{subsec:RQ2}). Our results also showed that \method generalizes across all the studied backends except for Rochester, Cambridge, Rainbow, Weber, 9q-square, and Toronto. In the future, we will conduct dedicated experiments to study whether \method can be generalized to these backends and beyond. In terms of quantum programs, \method generalizes for programs with similar output distributions. Nevertheless, it is important to note that all approaches have concerns regarding generalizability, and there is no guarantee that \method will generalize to other backends that are not part of IBM, Google, and Rigetti. We tried to utilize backends from the three leading companies in the field, considering the constraints of conducting a feasible experiment within a limited time frame.}

\section{Discussions} \label{sec:overalldiscussion}
\subsection{Generalizability}
\textbf{\textit{Generalizability of \method's modules.}}\label{sec:genreral_module} 
\method is designed modularly to incorporate any new backend without a major change to the module structure. Specifically, \method expects a baseline circuit as a QASM file, a common format Qiskit uses. To enhance the modularity, \method provides an abstract interface class, which can be used to define the logic for executing a QASM circuit on any noisy backend. Given the baseline circuits, \method automatically generates the required dataset for \textit{MLP Training} and uses an automated selection of optimal hyperparameters for the generated dataset. \revision{In terms of \textit{Feature Generation}, each quantum gate is affected by a certain type of noise, and different gate combinations accumulate, leading to different noise patterns, consequently affecting the programs' outputs. Hence, to generalize for covering different types of noises, we extract training features from the outputs of a quantum program; otherwise, the feature extraction would become gate- and circuit-specific. These features capture the accumulated effect of noise without needing to understand the details of noise patterns of specific gates and how they are combined to form quantum circuits.} In terms of \textit{Test Analyzer}, since \method only requires the output of a quantum program to generate features for \textit{Noise Filter}, \method can be integrated with any test strategy that assumes the availability of program specifications in the form of program outputs by acting as a filter between test execution and test assertion.

\textit{\textbf{Generalizability of the machine learning (ML) models of \method.}}
In the context of ML, generalizability refers to the ability of a model to perform well on unseen or new data that are different from those the model was trained on. In other words, generalizability indicates how well the model can learn and apply its knowledge to new situations. However, quantum noise is not only computer-specific but also circuit-specific. The general noise pattern of each NISQ computer is different~\cite{Martina2022}, which makes the learning of noise patterns specific to each NISQ computer. Developing a (generalized) single model covering all possible NISQ computers requires a further understanding of the characteristics of the NISQ computers and their quantum noise. For now, it is challenging to achieve because each NISQ computer has different noise characteristics directly correlating with the hardware configuration, physical implementation of qubits, and the surrounding environment~\cite{noise_benchmark1}. Moreover, the same quantum circuit exhibits a completely different noise distribution on different NISQ computers based on gate composition and dynamic circuit mapping on physical hardware~\cite{noise_benchmark2}. Currently, no set of generalized features can accurately define or map the behavior of quantum noise for all NISQ computers and circuits. However, new studies show that Bayesian inference might be a possible alternative to characterize noise for different NISQ computers and circuits~\cite{Bayesian2, Bayesian3}, which we will investigate in the future.

\subsection{Applicability and Maintainability}
One key aspect that hinders the maintainability aspect of \method is the concept of data drift. Data drift refers to the phenomenon where the statistical properties of the input data change over time, leading to a degradation in the performance of ML models~\cite{drift1}. Handling data drift is crucial to maintaining the accuracy and effectiveness of the models in real-world scenarios. The behavior of noise for a NISQ computer changes more frequently than any other data over time. To keep the noise model updated, IBM recalibrates its noise models every 24 hours\footnote{\url{https://quantum-computing.ibm.com/admin/docs/admin/calibration-jobs}}. Although handling data drift is not in the current scope of \method, however \method can be easily integrated with the current best strategies for handling data drifts such as adaptive learning~\cite{adaptive1}, data drift mitigation~\cite{drift1} and detection~\cite{drift2}. 

From the application perspective, a tester shall perform three steps to test a quantum program (i.e., $p$) on a noisy backend (i.e., $b$) with a test suite (i.e., $ts$) generated with any test method: (1) The tester obtains the baseline model of $b$ from our repository if available; otherwise, the tester can use the \textit{Baseline Trainer} component for training a baseline model from scratch; (2) The tester then tunes the baseline model for $p$ with a subset of inputs from $ts$ using the \textit{Baseline Tuner} component provided by \method; and (3) the tester uses the \textit{Test Analyzer} component provided by \method to test $p$ against $ts$, including filtering of results. 

When it comes to the applicability of \method as a whole, one might wonder why \method is needed at all since quantum programs can be executed on noise-free simulators. It is needed because quantum simulators are computationally expensive; even a small quantum program can take up to 12 hours to execute one input~\cite{simulators_are_slow}. As a result, only small quantum circuits can be executed and tested on simulators. Though, in the literature, some strategies (e.g., CutQC~\cite{cutqc}) have been proposed to reduce the computational cost, testing is still time-consuming. \method solves this problem by allowing testing directly on NISQ devices. \textit{Baseline Trainer} of \method requires simple quantum circuits without high execution costs to capture the general noise pattern of a backend, as shown in RQ1. The program specification for such circuits can be generated by simulation or can be provided by developers. \textit{Baseline Tuner} of \method requires only a handful of inputs to generate a circuit-specific model. For CUTs, if the program specification is unavailable, it can be generated using strategies like CutQC~\cite{cutqc} on noise-free simulators. The time cost of simulating only a handful of inputs is way less than doing systematic testing on noise-free simulators. By generating the circuit-specific model, the whole testing can be performed on a NISQ computer, which is much more scalable.

\revision{\subsection{Data requirements}
ML algorithms often require a significant amount of data, and the volume needed depends on the complexity of the problem at hand and the size of the ML model~\cite{datarequired}. However, in the context of quantum programs, generating a large dataset can be particularly time-consuming. As mentioned in Sect.~\ref{introduction}, simulating a quantum program for a single input can be a time-intensive process. Additionally, quantum noise is unique to the backend on which a quantum circuit is executed. Training a model for all possible quantum circuit-backend pairs would demand massive data and necessitate larger, more complex ML models, increasing the training and inference time costs.}
\revision{In software test assessment, testing activities often face budget constraints, and quantum programs pose a significant challenge due to their vast state space. To sufficiently test a quantum program, a substantial number of test input executions are required. Given these limitations, employing a complex ML model that demands extensive data and large inference times may not be a practical solution. Having a balance between model complexity and the practical considerations of testing resources is crucial.}

\revision{To address data-related challenges, we have divided our approach into two distinct modules: Baseline Trainer and Baseline Tuner. Baseline Trainer primarily focuses on learning general noise patterns. Therefore, the data requirements for Baseline Trainer involve collecting observed noisy quantum output states from multiple quantum circuits, which is a one-time cost for each NISQ computer. In contrast, Baseline Tuner specializes in learning circuit-specific noise patterns. Consequently, the data requirements for Baseline Tuner consist of the noisy output states from a specific circuit. This division of the problem into two stages reduces the required data and allows for the use of less complex ML models. In our experiments, the amount of generated data depends on the number of selected baseline circuits and the number of output states produced by each circuit-input pair. For a real-program experiment with only three baseline circuits, the generated data for ML consists of approximately 1000 output states for each backend. For the artificial program experiment with 200 baseline circuits, the generated data exceeds 5000 output states. For both experiments, we chose a multi-layer perceptron (MLP) model with only two hidden layers, as increasing the number of layers would result in overfitting due to the limited amount of data available. Our experimental findings show that \method delivers satisfactory results for most backends. However, it becomes evident that additional training data or a more complex model is needed for certain backends where improvements are limited. Nevertheless, obtaining more training data through simulation would significantly increase training costs. Consequently, the amount of data required highly depends on the particular quantum computer targeted.}


\revision{\subsection{Time cost}
The adoption of \method enhances the reliability of test assessments on NISQ computers, but it does come at the expense of increased time required for the overall test assessment of a quantum circuit. Here, we discuss the time cost associated with applying \method in relation to its core components. First, the Data Generation component within the Baseline Trainer module generates the necessary data to train a baseline MLP model for a specific noise backend. Its time cost is a one-time expense for each noisy backend and is directly influenced by the number of baseline circuits and the quantum simulator utilized. Our experiments had three baseline circuits for experiment one and 200 baseline circuits for experiment two. The average time cost of the Data Generation component for one noisy backend amounted to 11.8 minutes. For a real quantum computer, the time cost would typically be in the order of seconds. Second, the MLP training component of Baseline Trainer also has a one-time cost for a given noisy backend, which is, on average, approximately one minute for our experiments conducted on an Nvidia 3080 GPU. This highlights that although the time cost for the Baseline Trainer module is incurred only once, the quantum simulator occupies a significant amount of time compared to model training due to the current hardware limitations.}

\revision{There is also a time cost associated with the Baseline Tuner module, which is not a one-time cost since it depends on how frequently noise changes in the noisy backend. Typically, for IBM quantum computers, the noise models are calibrated at least once every 24 hours, implying that the Baseline Tuner module needs to be executed once every 24 hours. In our experiments, the average time cost for the Baseline Tuner module for one circuit-backend pair was approximately 14 minutes. About 12 minutes were required for data generation to fine-tune a baseline MLP model. Note that the data generation cost would significantly decrease when performed on NISQ computers. This is attributed to the fact that the execution of a quantum program on a NISQ computer typically takes seconds, whereas, on simulators, it takes several minutes. This indicates that the MLP tuning cost is roughly 2 minutes for one circuit-backend pair.}

\revision{So, how much additional time is needed for test assessment when \method is integrated? The time cost of test assessment of a single circuit-backend pair would be the time cost of the Baseline Tuner and the inference time required by the Noise Filter component. The inference time cost for the MLP model on a GPU is only a few seconds. This underscores that \method does not entail a substantial time cost for the test assessment of a single circuit-backend pair.}

\subsection{Limitations}
\revision{First, \method only provides the baseline models of 26 noisy backends (23 from IBM, two from Google, and one from Rigetti). Therefore, a user can use \method to test quantum programs on these 26 noisy backends with any test strategy that uses program outputs and their probabilities to check the correctness of quantum programs.} As a result, additional effort is needed to learn baseline models for other backends. This would require having the noisy outputs of the baseline circuits to train a new baseline model for a new noisy backend (see Generalizability of modules in Sect.~\ref{sec:genreral_module}). 

Second, noisy backends keep changing; therefore, their corresponding baseline models need to be updated regularly. To this end, \method is expected to benefit from adaptive learning and online learning~\cite{adaptive1, onlinelearning}. Third, 
we need to conduct research on integrating different models as one model, such that it could model the noise of all backends or, at least, a subset of backends with similar characteristics (e.g., number of qubits, qubit layouts). One possible aspect could be the adoption of Bayesian inference models~\cite{Bayesian} for this purpose. Lastly, we only explored using the program output diversity as the criterion to generate diverse quantum programs. However, other criteria also deserve attention, such as structural diversity, circuit depth, or combinations of multiple diversity criteria.


\section{Conclusion and Future Work}\label{sec:conclusion}
To enable quantum software testing techniques to deal with inherent hardware noise--inevitable in current and near-term quantum computers, we presented an approach called \method. The approach applies ML techniques to learn hardware and quantum circuit-specific noise, followed by filtering noise effects from program outputs. As a result, filtered outputs are used for test case assessment against a given test oracle, which is much more accurate than test case assessment of unfiltered outputs. With \method, quantum software testers can determine whether a test case failed due to real faults or noise. We evaluated \method using nine real quantum programs and 1000 diverse quantum programs generated with IBM's Qiskit framework.
Moreover, faulty versions of these programs were also generated to determine whether \method allows to correctly detect failing test cases, i.e., whether it allows for distinguishing between faults and noise.
\revision{We used 23 noise models from IBM, two available noise models from Google, and one noise model from Rigetti to run experiments. Our results showed that \method could effectively remove noise, thereby providing testers with a tool to determine whether a quantum program has a real fault.}

\revision{In the future, we will extend our experiments to larger sets of quantum programs with various criteria for measuring diversity. Moreover, we also plan to verify \method on real quantum computers instead of simulators. Furthermore, we plan to integrate \method with other quantum software testing techniques.}

\bibliographystyle{IEEEtran}

\end{document}